\documentclass[11pt]{article}
\usepackage[utf8]{inputenc}
\usepackage[left=2cm,right=2cm,top=2cm,bottom=2cm]{geometry}
\usepackage{amsmath}
\usepackage{amsfonts}
\usepackage{amssymb}
\usepackage{xcolor}
\usepackage{slashed}
\usepackage{graphicx}
\usepackage{multirow}
\usepackage[bf]{caption}
\usepackage[numbers,sort&compress]{natbib}
\usepackage{enumerate}
\usepackage{hyperref}
\usepackage{dsfont}
\def\dx{\mathrm{d}x}
\def\dy{\mathrm{d}y}
\def\dz{\mathrm{d}z}
\def\mW{m_W}
\def\mnui{m_{\bf i}}
\def\mnuj{m_{\bf f}}

\newcommand{\ini}{{\bf i}}
\newcommand{\fin}{{\bf f}}

\DeclareMathOperator\PL{P_L}
\DeclareMathOperator\PR{P_R}

\newcommand{\be}{\begin{equation}}
\newcommand{\ee}{\end{equation}}
\newcommand{\beq}{\begin{equation*}}
\newcommand{\eeq}{\end{equation*}}

\allowdisplaybreaks

\providecommand{\href}[2]{#2}
\begin{document}
\begin{center}
{\Large\bf Asymmetry in flavour changing electromagnetic transitions of vector--like quarks}
\end{center} 
\vspace{0.2cm}

\begin{center}
{\bf Shyam Balaji}\footnote{\tt sbalaji@lpthe.jussieu.fr}\footnote{\tt shyam.balaji@sydney.edu.au}
\\\vspace{5mm}
{$^1$LPTHE, UMR 7589 Sorbonne Université \& CNRS, 4 Place Jussieu, F-75252, Paris, France} \newline
{$^2$School of Physics, The  University of Sydney, NSW 2006, Australia} \\
\end{center}

\vspace{1.5cm} 

\begin{abstract} 
Vector--like quarks have been of interest for a plethora of experimentally motivated reasons and have come under increased investigation recently due to their inclusion to the Standard Model (SM) appreciably improving global fits to several flavour physics and precision electroweak measurements. The addition of vector--like quark singlets breaks the unitarity of the Cabibbo-Kobayashi-Maskawa (CKM) matrix and enables tree--level flavour changing neutral (FCN) vertices. The resulting radiative flavour changing decays of these particles through the electromagnetic transition dipole moment are a key means to study their properties and search for them at experiments like the Large Hadron Collider (LHC). Despite these radiative decays providing such a clean experimental signature, an explicit analytical study of the branching ratios and $CP$ violation resulting from these loop level processes has thus far evaded attention. We provide the formulation for the decay rates and $CP$ asymmetry resulting from a combination of the imaginary components of the loop integrals and complex phases in the quark mixing matrix. We then apply our analytical results to study phenomenology of these states for several global fits pertaining to vector--like isosinglets $t'$ and $b'$. We find that clean collider signatures and polarisation observables can be generated for both $t'$ and $b'$.
\end{abstract}
\begin{flushleft}
\hspace{0.8cm} PACS number(s): \\
\hspace{0.8cm} Keywords: Heavy quark, $CP$ violation, radiative decay, flavour physics, polarisation
\end{flushleft}

\def\thefootnote{\arabic{footnote}}
\setcounter{footnote}{0}

\newpage

\section{Introduction}

The study of radiative particle decays has been of interest for many decades because they provide a potential experimentally clean probe for new physics \cite{Deshpande:1981zq, Beneke:2000hk}. The electromagnetic dipole moment of heavy quarks can be generated at various loop levels and they induce radiative decays when off--diagonal components of the dipole moment are nonzero. This is analogous to the flavour changing radiative decays of the top quark \cite{Balaji:2019fxd, Balaji:2020fxd, Balaji:2020qjg}. Precision measurements and searches for flavour changing decays of heavy quarks through the  electromagnetic transition dipole moment provide an alluring and distinct signature for new physics beyond the Standard Model (SM) \cite{Balaji:2020fxd}.

The SM contains three generations of up and down--type quarks. All quarks with a given electric charge mix, hence there is a charged current coupling between each down-- and up--type quark. These couplings are summarised in the Cabibbo--Kobayashi--Maskawa (CKM) matrix. However, there is no known constraint in nature requiring only three generations of quarks and therefore the existence of exotic quarks has long been hypothesised. Such states may include heavy vector--like quarks (VLQs), these states do not exist in the SM, but they naturally arise near the electroweak scale in many new physics models including the minimal super--symmetric model, Left--Right symmetry models, top colour assisted technicolour and two Higgs doublets with four generations of quarks \cite{Dedes:2014asa, Hill:1994hp, Gaitan:2015hga, Atwood:1996vj,Arhrib:2005nx,DiazCruz:1989ub,Atwood:1995ud, Atwood:1995ej}. Flavour hierarchies among SM fermion masses and their mixing are most often generated through their dynamical mixing with such vector--like fermions \cite{Balaji:2018zna,Balaji:2019kwe}.

The possibility of gauging the SM flavour group requiring anomaly cancellation via addition of new vector--like fermions has been studied explicitly and the possibility of a relatively low new gauge boson scale emerging was shown \cite{Grinstein:2010ve}. In such models, VLQs play a crucial role in both anomaly cancellation and fermion mass generation. The VLQs themselves also mix with the lighter generations in phenomenologically interesting ways. The simplest VLQ is an isosinglet with both left-- and right--handed components transforming as singlets under $SU(2)_L$ with $I_3=0$ that can either be up-- or down--type which are labelled $t'$ and $b'$ respectively.\footnote{Often the labels T and B are used for vector--like quarks $t'$ and $b'$ in the literature. These labels are used to explicitly denote the new particles as top and bottom partners respectively.}

The Large Hadron Collider (LHC) is currently collecting data that will enable testing and potential discovery of this sector in the near future. Typical cross--sections for single and pair VLQ production are fairly large and within short term reach for the LHC detectors \cite{Cacciapaglia:2010vn}. Although limits on mass scales of $m_{t'}>1.31$ TeV and $m_{b'}>1.22$ TeV at the 95\% C.L.~\cite{ATLAS:2018ziw} have been placed at the ATLAS detector, no exotic quark signatures have thus far been observed at the LHC. Alternatively, indirect signals of VLQs can be considered due to the loop--level contributions of the exotic quarks in SM particle processes. Precision flavour measurements place strong limits on the new heavy quarks and set the lowest mass scale and maximum mixing for these states.

In this work, we will focus on a precise computation of the branching ratios for the radiative VLQ decays based on extended CKM matrix best fit values. We focus particularly on the computation of the $CP$ asymmetry that arises due to the complex phase generated by the loop integrals, this implies asymmetric decay rates $\Gamma(q_\ini \rightarrow q_\fin\gamma) \neq \Gamma(\bar{q}_\ini\rightarrow \bar{q}_\fin\gamma)$. Such a $CP$ asymmetry generates circular polarisation between the two circularly polarised photons $\gamma_{+}$ and $\gamma_{-}$ \cite{Boehm:2017nrl,Balaji:2019fxd} and hence provides a crucial and experimental clean probe for new physics. We provide the analytical formulation for the loop functions for each process and the $CP$ asymmetry ratios in terms of chiral form factors which characterise the  electromagnetic transition dipole moment. Although radiative decays of VLQs have previously been considered \cite{Cacciapaglia:2010vn}, detailed theoretical and phenomenological studies of the resulting polarisation observables have evaded attention. It has been shown previously that SM radiative decay channels of the top quark are currently unobservable due to the Glashow-Iliopoulos-Maiani (GIM) suppression \cite{Eilam:1990zc}. We find that branching ratios for the VLQs studied here are enhanced by several orders of magnitude since including them breaks the GIM mechanism and enables flavour changing neutral (FCN) vertices at tree--level \cite{AguilarSaavedra:2002ns, AguilarSaavedra:2002kr}. This means that the $3\times3$ quark mixing submatrix that corresponds to the CKM matrix is no longer necessarily unitary. 


The outline of the paper is as follows, we first derive the VLQ electromagnetic transition dipole moment in Section~\ref{sec:vlqemdm}. We divide this section further into an overview of the interaction Lagrangian, computation of all the relevant Lorentz invariant amplitudes, analytical evaluation of the loop integrals and chiral form factors followed by calculating the resulting $CP$ asymmetries for radiative decays. This is accompanied by the results in Section~\ref{sec:results} which contains several phenomenological studies. In Section~\ref{sec:CPuptype}, we consider an up--type VLQ isosinglet $t'$. The section is further divided into Section~\ref{sec:newsignalsuptype}, where we consider a $4\times3$ quark mixing matrix that is strongly constrained by flavour physics measurements from Ref.~\cite{Alok:2015vvk} and Section~\ref{sec:CKMunitarityproblem}, where recently observed hints of unitarity deviation in the first row of the CKM matrix is explained by inclusion of a $t'$ \cite{Belfatto:2019swo, Belfatto:2021jhf, Cacciapaglia:2010vn}. In Section~\ref{sec:CPdowntype} we instead consider a $3\times4$ quark mixing matrix with down--type VLQ isosinglet $b'$. The strong constraints on $b'$ are derived from Ref.~\cite{Alok:2014yua}. In all fits considered, we find compelling phenomenology for the VLQ and discuss implications for exotic quark searches.

\section{Calculating form factors and \texorpdfstring{$CP$}{TEXT} violation}
\label{sec:vlqemdm}
\subsection{Interaction Lagrangian}
\label{sec:lagrangian}
The charged current interaction Lagrangian in the mass eigenstate basis needed for the decay channels of interest is given by
\begin{equation}
\mathcal{L}_W  = -\frac{g}{\sqrt{2}}\bar{u}_\alpha\gamma^\mu\PL V_{\alpha \beta}d_\beta W_\mu^{+}+ \mathrm{h.c.},
\label{eq:LagrangianW}
\end{equation}
where $\alpha=1,...,n_u$ and $\beta=1,...,n_d$, the usual chiral projection operators are defined as $\mathrm{P_{\text{L,R}}} = \frac{1}{2}(1 \mp \gamma_5)$ and $V$ is the quark mixing matrix. In the SM, $n_u=n_d=3$ where we have flavours $u_i=(u,c,t)$ and $d_i=(d,s,b)$ and therefore $V$ reduces to the $3\times3$ CKM matrix. In this work we will be interested in extensions where $(n_u,n_d)=(4,3)$ as shown in Section~\ref{sec:CPuptype} and $(n_u,n_d)=(3,4)$ as shown in Section~\ref{sec:CPdowntype}, which include a new vector isosinglet $t'$ and $b'$ respectively. For neutral current interactions we get  
\begin{equation}
\mathcal{L}_Z  = -\frac{g}{2 \cos\theta_w}\bar{q}_\alpha\gamma^\mu (c_L\PL + c_R\PR) q_\beta Z_\mu + \mathrm{h.c.},
\label{eq:LagrangianZ}
\end{equation}
where the chiral component coefficients are given by $c_L=\pm X^q_{\alpha\beta}-2Q_q\sin^2\theta_w\delta_{\alpha\beta}$ and $c_R=-2Q_q\sin^2\theta_w \delta_{\alpha\beta}$ and $q=u_i,d_i$. If $q$ is an up--type quark the $c_L$ off--diagonal component is taken to be positive while for a down--type it is taken to be negative and the quark electric charges are given by $Q_u=2/3$ and $Q_d=-1/3$. It should also be noted that the Hermitian matrices $X^u$ and $X^d$ are defined
\begin{align}
    X^u_{\alpha\beta}&= (V V^\dagger)_{\alpha\beta}\,,&X^d_{\alpha\beta}&= ( V^\dagger V)_{\alpha\beta}.
\end{align}
The quark interactions with the Goldstone charged scalar $\phi^\pm$, which corresponds to the longitudinal mode of the $W$ boson, is given by
\begin{align}
 \mathcal{L}_{\phi}=-\frac{g}{\sqrt{2} m_W}\bar{u}_\alpha V_{\alpha\beta}(m_\alpha  \PL- m_\beta \PR)d_\beta\phi^+ +\mathrm{h.c}.
\end{align}
The Lagrangian for the neutral Goldstone scalar $\chi$, this time representing the longitudinal mode of the $Z$ boson is given by
\begin{align}
     \mathcal{L}_{\chi}=\frac{i g}{2 m_W}\left[\bar{u}_\alpha X^u_{\alpha\beta}(m_\alpha  \PL-m_\beta \PR)u_\beta-\bar{d}_\alpha X^d_{\alpha\beta}(m_\alpha  \PL-m_\beta \PR)d_\beta \right]\chi +\mathrm{h.c}.
\end{align}
Finally the interaction of the quarks with the Higgs boson $h$ is given by
\begin{align}
    \mathcal{L}_h=\frac{g}{2m_W}\left[\bar{u}_\alpha X^u_{\alpha\beta}(m_\alpha  \PL+m_\beta \PR)u_\beta+\bar{d}_\alpha X^d_{\alpha\beta}(m_\alpha  \PL+m_\beta \PR)d_\beta \right]+\mathrm{h.c}.
\end{align}
From here, we may derive the Feynman rules and calculate amplitudes for the decay processes of interest.
\subsection{Analytical derivation of amplitudes and form factors}

\begin{figure}[h!]

\hspace{-1.6cm}\includegraphics[width=1.2\linewidth]{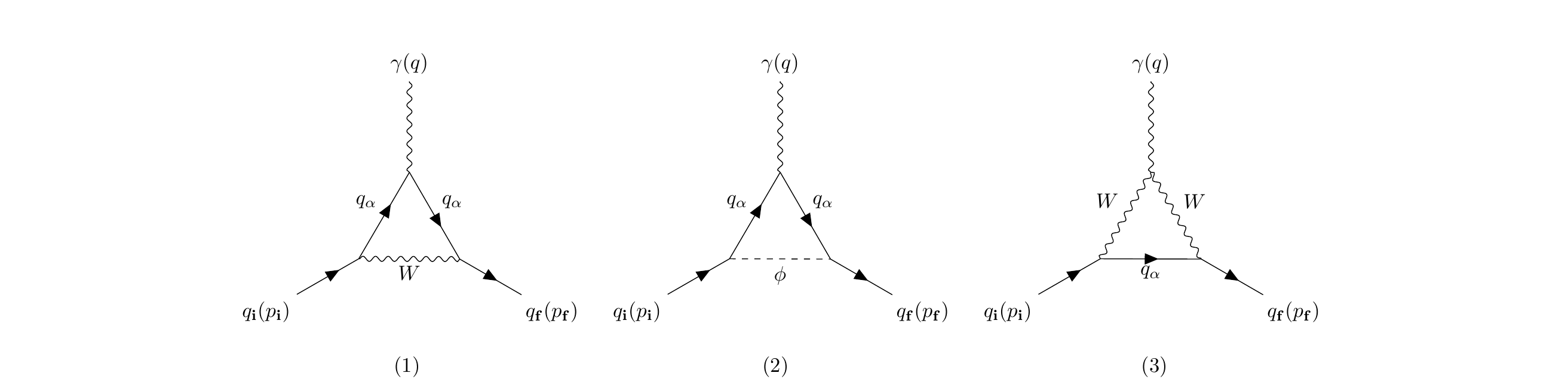}

  \hspace{-1.6cm}\includegraphics[width=1.2\linewidth]{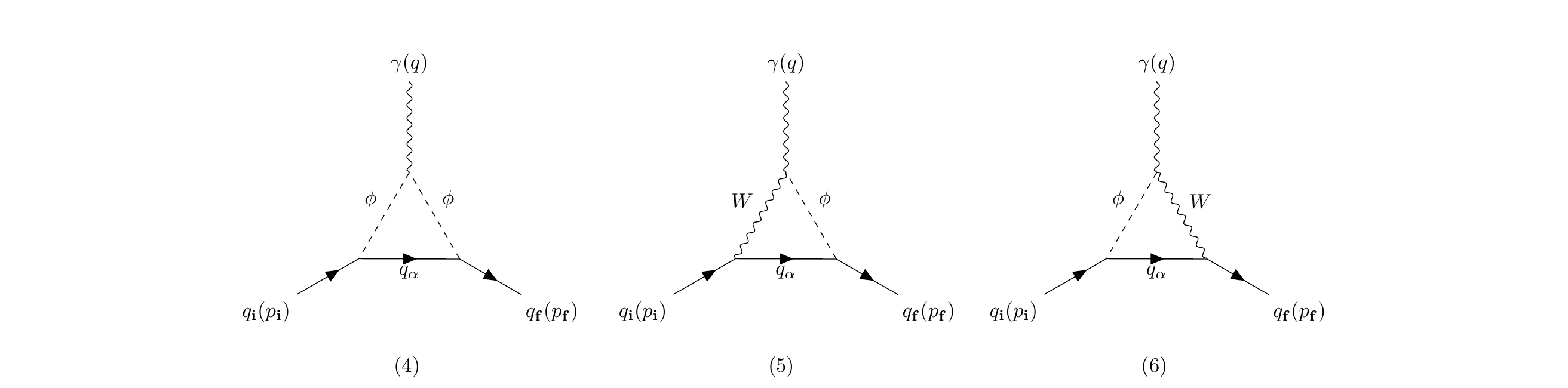}
   
  \hspace{-1.6cm} \includegraphics[width=1.2\linewidth]{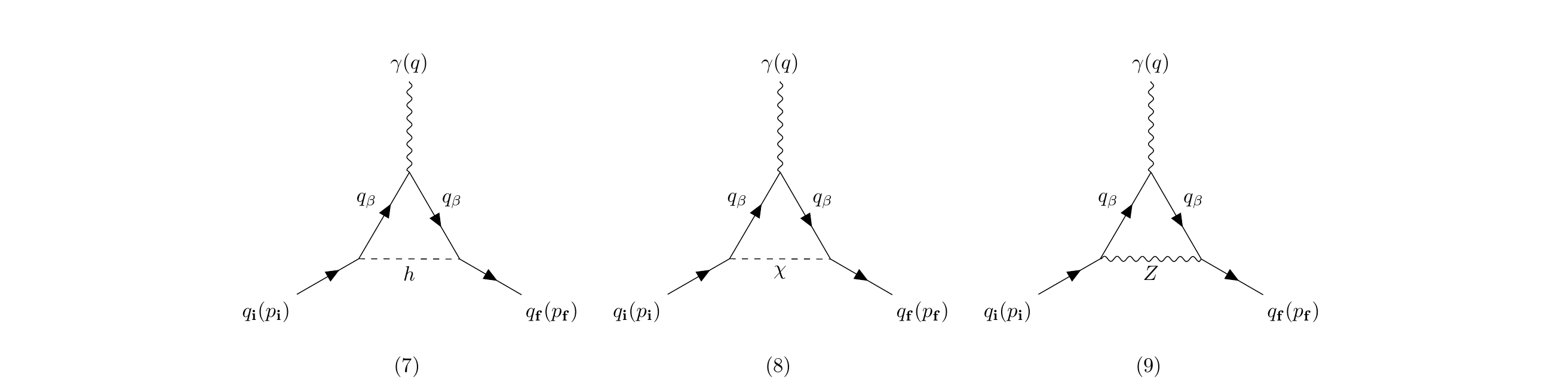}

    \caption{Feynman Diagrams for the one-loop  vector--like quark decay induced by electroweak interactions. We denote the amplitudes for the nine diagrams as $\mathcal{M}_1-\mathcal{M}_9$ accordingly. We note that amplitudes $\mathcal{M}_7-\mathcal{M}_9$ contain flavour changing neutral vertices. We denote $q=p_\fin-p_\ini$ and $p$ is the loop momentum running through the bottom edge of each diagram that is integrated over. }\label{fig:SM_loops}
\end{figure}

The unique contributions to the vector--like quark decay through the transition dipole moment are shown in Fig.\ref{fig:SM_loops}. 

The resulting matrix elements can be written in terms of a generic vertex function $\Gamma^\mu_{\fin \ini }$ like 
\begin{eqnarray}
i \mathcal{M} (q_\ini \to q_\fin + \gamma_{\pm})  =  \bar{u}(p_\fin) \Gamma_{\fin \ini }^\mu(q^2) u(p_\ini) \varepsilon^*_{\pm,\mu}(q) \,. \label{eq:decay_amplitude}
\end{eqnarray}
By requiring the photon (or gluon) to be on-shell $q^2=0$ and choosing the Lorenz gauge $q\cdot \varepsilon_p = 0$, the anapole does not contribute to $\Gamma^\mu_{\fin \ini }$. Therefore, the vertex function is simplified to \cite{Balaji:2019fxd}
\begin{eqnarray} \label{eq:dipole}
\Gamma^\mu_{\fin \ini }(q^2) &=& -f^{\rm{M}}_{\fin \ini}  (i \sigma^{\mu \nu} q_\nu) + f^{\rm{E}}_{\fin \ini}  (i \sigma^{\mu \nu} q_\nu \gamma_5) \,, 
\end{eqnarray}
where $f^{\rm{E}}_{\fin \ini} $ and $f^{\rm{M}}_{\fin \ini}$ are the electric and magnetic transition dipole moments of $q_\ini \to q_\fin \gamma$ respectively.  We may separate the electromagnetic transition dipole moment into chiral components like 
\begin{eqnarray}
\label{eq:vertex_function}
\Gamma^\mu_{\fin \ini } (q^2) &=& i \sigma^{\mu \nu} q_\nu [f^\text{L}_{\fin \ini } (q^2) \PL + f^\text{R}_{\fin \ini } (q^2) \PR] \,, 
\end{eqnarray}
where $f^\text{L}_{\fin \ini }$ and $f^\text{R}_{\fin \ini }$ are the left and right chiral form factors. They satisfy the relations $f^{\text{L,R}}_{\fin \ini} = -f^{\rm{M}}_{\fin \ini} \pm i f^{\rm{E}}_{\fin \ini} $ \cite{Balaji:2019fxd}.

For each diagram in Fig.~\ref{fig:SM_loops}, we may explicitly write down the Lorentz invariant amplitudes in Appendix~\ref{sec:feynmanamplitudes} using the 't Hooft--Feynman gauge, where we utilise the following definitions in the case of an initial state up--type quark
\begin{align}
\label{eq:uptypecoefficients}
    A_\alpha&= \frac{e g^{2}}{2} V^*_{\ini \alpha } V_{\fin\alpha }\,,  &C_\beta&= \frac{e |Q_u| g^2}{4} X^u_{\fin\beta } X^u_{\beta \ini} \,, \nonumber \\
         C^L_{\mu\nu} &=\frac{e |Q_u| g^2}{4 \cos^2{\theta_w}}[ X^u_{\mu\nu}-2Q_\nu\sin^2\theta_w\delta_{\mu\nu} ]\,,  &C^R_{\mu\nu} &=-\frac{e |Q_u| Q_\nu g^2 }{2}\tan^2\theta_w \delta_{\mu\nu}\,, 
\end{align}
where for coefficients $A$ and $C$ the indices $\alpha=d_i$ and $\beta=u_i$ while in the case of an initial state down--type quark we have
\begin{align}
\label{eq:downtypecoefficients}
    A_\alpha&= \frac{e g^{2}}{2} V_{ \alpha \ini } V^*_{\alpha \fin}, &C_\beta= \frac{e |Q_d| g^2}{4} X^d_{\fin\beta } X^d_{\beta \ini}\,,  \nonumber \\
        C^L_{\mu\nu} &=\frac{e |Q_d| g^2}{4 \cos^2{\theta_w}}[- X^d_{\mu\nu}-2Q_\nu\sin^2\theta_w\delta_{\mu\nu} ], & C^R_{\mu\nu} =-\frac{e |Q_d| Q_\nu g^2 }{2}\tan^2\theta_w \delta_{\mu\nu}\,, 
\end{align}
with $\alpha=u_i$ and $\beta=d_i$ for $A$ and $C$ respectively. It should be noted that for diagrams with charged currents and Goldstone modes $\mathcal{M}_1$ to $\mathcal{M}_{6}$, the index $\alpha$ denotes quarks in the loop. In the case of an up-- or down--type quark in the initial state, $\alpha$ is a down-- and up--type quark respectively. Conversely, for the diagrams with Higgs, neutral Goldstone modes and currents shown in $\mathcal{M}_7$ to $\mathcal{M}_9$, we instead use the index $\beta$ for the quarks running in the loop. In this case, for an up-- or down--type quark in the initial state, $\beta$ in the loop is an up-- or down--type quark respectively. This must be satisfied since all the bosons in the loop are electrically neutral (hence $Q_\beta=Q_\ini$ must be satisfied). If we want to consider the case where a gluon is radiated rather than a photon, we must make the replacement $e\to g_s$ and $Q_u, Q_d\to 1$ in Eqs.~\eqref{eq:uptypecoefficients} and \eqref{eq:downtypecoefficients}. 

We now follow the standard Feynman parametrisation procedure to integrate over loop momenta. We then apply the  Gordon decomposition to separate chiral components, and finally factorise electromagnetic transition dipole moment terms for the first six diagrams with coefficients as 

\begin{eqnarray}\label{eq:effective_vertex}
&\Gamma_{\fin \ini, \alpha}^{\mu,({\rm k})}&=
i \sigma^{\mu\nu}q_\nu \int_0^1\dx\dy\dz\,\delta(x+y+z-1)\, {\cal P}^{( k)} \,,
\end{eqnarray}
where for ${k}=1,..,6$ we get
\begin{eqnarray}\label{eq:effective_vertex1}
&{\cal P}^{(1)}&= \pm\frac{|Q_\alpha| A_\alpha}{16\pi^2}\frac{-2x (x+z) m_\ini \PR - 2x (x+y) m_\fin \PL}
{  m_\alpha^2(1-x)+x m_W^2-x (y m_\ini^2 + z m_\fin^2)}\,, \nonumber\\
&{\cal P}^{(2)}&=\pm\frac{|Q_\alpha| A_\alpha}{16\pi^2}\frac{[x z m_\fin^2 -((1 - x)^2 +xz) m_d^2] m_\ini \PR + [xy m_\ini^2 - ((1-x)^2 + xy) m_d^2] m_\fin \PL}
{ m_W^2 \left[ m_\alpha^2(1-x)+x m_W^2-x (y m_\ini^2 + z m_\fin^2)\right]}\,, \nonumber\\
&{\cal P}^{(3)}&= \frac{A_\alpha}{16\pi^2}\frac{[(1-2x)z-2(1-x)^2] m_\ini \PR + [(1-2x)y-2(1-x)^2] m_\fin \PL}
{m_W^2(1-x)+x m_\alpha^2-x (y m_\ini^2 + z m_\fin^2)} \,, \nonumber\\
&{\cal P}^{(4)}&= \frac{A_\alpha}{16\pi^2}\frac{[xz m_\fin^2 - x (x+z) m_\alpha^2] m_\ini \PR + [xy m_\ini^2 - x (x+y) m_d^2] m_\fin \PL}
{m_W^2\left[m_W^2(1-x)+x m_\alpha^2-x (y m_\ini^2 + z m_\fin^2)\right]} \,, \nonumber\\
&{\cal P}^{(5)}&= \frac{A_\alpha}{16\pi^2}\frac{-z m_\ini \PR} {m_W^2(1-x)+x m_\alpha^2-x (y m_\ini^2 + z m_\fin^2)} \,, \nonumber\\
&{\cal P}^{(6)}&= \frac{A_\alpha}{16\pi^2}\frac{-y m_\fin \PL} {m_W^2(1-x)+x m_\alpha^2-x (y m_\ini^2 + z m_\fin^2)} \,. 
\end{eqnarray}
We note the that the plus or minus signs for ${\cal P}^{(1)}$ and ${\cal P}^{(2)}$ for photon and gluon emission respectively follow from the corresponding matrix elements shown in Section~\ref{sec:feynmanamplitudes}. Since the diagrams with charged bosons in the loop have a similar form to those shown in Ref.~\cite{Balaji:2020fxd}, we may combine all contributions to the vertex function and consequently the right--handed form factors are given by
\begin{align}
\label{eq:SM_form_factors}
    {f}^{\rm R (c.c)}_{\fin \ini, \alpha} &=\sum_{k=1}^6 \Gamma_{\fin \ini, \alpha}^{\mu,({ k})} \PR, & {f}^{\rm R (g)}_{\fin \ini, \alpha} =\sum_{k=1}^2 \Gamma_{\fin \ini, \alpha}^{\mu,({ k})} \PR.
\end{align}
Note the left--handed form factors are analogous except the projection operator chirality is switched. In the above expression, the label $l$ can refer to photon ($\rm c.c$) or gluon emission ($g$) and can be written in terms of the analytical loop functions as
\begin{align}
{f}^{\mathrm{R} (l)}_{\fin \ini, \alpha}&=\frac{A_\alpha}{16\pi^2}K_{\fin\ini}^{(l)}, &{f}^{\mathrm{R} (l)}_{ \ini\fin, \alpha}&=\frac{A_\alpha^*}{16\pi^2}K_{\ini\fin}^{(l)}, &{f}^{\mathrm{L} (l)}_{\fin \ini, \alpha}&=\frac{A_\alpha}{16\pi^2}K_{\ini\fin}^{(l)}, &{f}^{\mathrm{L} (l)}_{\ini\fin , \alpha}&=\frac{A_\alpha^*}{16\pi^2}K_{\fin \ini}^{(l)},\label{eq:SM_form_factor_relations}
\end{align}
where we have
\begin{align}
    K_{\fin\ini}^{(c.c)}&=\frac{m_\ini}{m_W^2}\Bigg\{\frac{ (1+|Q_\alpha|)(\Delta_{\alpha\fin}+2m_W^2)}{2\Delta_{\fin\ini} }+\int_0^1\dx \left[\frac{|Q_\alpha|\left((\Delta_{\ini\alpha}-2m_W^2)(m_\alpha^2+m_\fin^2x^2)+x m^4_{\fin\ini,\alpha}\right)}{\Delta_{\fin\ini}^2x}\log \Lambda_{\ini\fin\alpha W}\right.\nonumber\\[.5cm]
&\hspace{2cm}\left.+\frac{(\Delta_{\ini\alpha}-2m_W^2)(m_\alpha^2+m_\fin^2(x-1)^2)+(1-x)m^4_{\fin\ini,\alpha}}{\Delta_{\fin\ini}^2x}\log \Lambda_{\ini\fin W\alpha}\right]\Bigg\},\nonumber\\
    K_{\fin\ini}^{(g)}&=-\frac{m_\ini }{ m_W^2}\left\{\frac{ \Delta_{\alpha \fin}+2m_W^2}{\Delta_{\fin\ini}}+\int_0^1\dx \frac{(\Delta_{\ini\alpha}-2m_W^2)(m_\alpha^2+m_\fin^2x^2)+x m^4_{\fin\ini,\alpha}}{\Delta_{\fin\ini}^2x}\log\Lambda_{\ini\fin \alpha W}\right\},
\end{align}
and we use the definition
\begin{equation}
   m^4_{\fin\ini,\alpha}=2m_W^2m_\alpha^2-(m_\alpha^2+m_\fin^2-2m_W^2)(m_\ini^2-m_\alpha^2-m_W^2).
\end{equation}
We also invoke convenient notation for the squared mass difference $\Delta$ and ratio $\Lambda$ like 
\begin{align}
    \Delta_{ij}&=m_i^2-m_j^2, &\Lambda_{ijkl}&=\frac{m_k^2+(m_l^2-m_k^2-m_i^2)x+m_i^2x^2}{m_k^2+(m_l^2-m_k^2-m_j^2)x+m_j^2x^2},
\end{align}
for a concise analytical representation of the form factors. It should be noted that while replacing the order of indices for the functions $\Delta$, $\Lambda$ and $K$ simply requires exchanging masses, exchanged indices for the form factors is a nontrivial operation requiring conjugation of the coefficient terms \cite{Balaji:2019fxd}.

Repeating a similar process for the diagrams 7--9 in Fig.~\ref{fig:SM_loops}, with neutral bosons running in the loop and FCN vertices, we get
\begin{eqnarray}\label{eq:effective_vertex2}
&{\cal P}^{(7)}&=\frac{C_\beta}{16\pi^2} \frac{m_\ini(-x z m_\fin^2+(x(x+z)-1)m_\beta^2) \PR + m_\fin(-x y m_\ini^2+(x(x+y)-1)m_\beta^2) \PL}
{ m_W^2 \left[ m_\beta^2(1-x)+x m_h^2-x (y m_\ini^2 + z m_\fin^2)\right]} \,, \nonumber\\
&{\cal P}^{(8)}&= -\frac{C_\beta}{16\pi^2} \frac{[x z m_\fin^2 -((1 - x)^2 +x z) m_\beta^2] m_\ini \PR + [xy m_\ini^2 - ((1-x)^2 + xy) m_\beta^2] m_\fin \PL}
{ m_W^2 \left[m_\beta^2(1-x)+x m_Z^2-x (y m_\ini^2 + z m_\fin^2)\right]} \,, \nonumber\\
&{\cal P}_{LL}^{(9)}&= -\frac{C^L_{\fin\beta}C^L_{\beta\ini}}{16\pi^2}\frac{-2  x(x+z) m_\ini \PR -2  x(x+y) m_\fin \PL}
{m_\beta^2(1-x)+x m_Z^2-x (y m_\ini^2 + z m_\fin^2)}\,, \nonumber\\
&{\cal P}_{LR}^{(9)}&=-\frac{C^L_{\fin\beta}C^R_{\beta\ini}}{16\pi^2} \frac{ 4x m_\beta \PR}
{ m_\beta^2(1-x)+x m_Z^2-x (y m_\ini^2 + z m_\fin^2)}\,, \nonumber\\
&{\cal P}_{RL}^{(9)}&= -\frac{C^R_{\fin\beta}C^L_{\beta\ini}}{16\pi^2}\frac{4 x m_\beta \PL} {m_\beta^2(1-x)+x m_Z^2-x (y m_\ini^2 + z m_\fin^2)}\,, \nonumber\\
&{\cal P}_{RR}^{(9)}&= -\frac{C^R_{\fin\beta}C^R_{\beta\ini}}{16\pi^2}\frac{-2 x(x+y) m_\fin  \PR -2 x(x+z)m_\ini  \PL} {m_\beta^2(1-x)+x m_Z^2-x (y m_\ini^2 + z m_\fin^2)}\,. 
\end{eqnarray}

Now we may integrate over the Feynman parameters $y$ and $z$ and complete our derivation of the chiral form factors for the FCN vertex diagrams. These are shown in full detail in Appendix~\ref{sec:formfactorrelations} where we have utilised the definitions

\begin{align}
\label{eq:fcnloopfunctions}
    K_{\fin\ini}^{(7)}&=\int dx \frac{m_\ini}{\Delta_{\fin\ini}^2 m_W^2 x}\left[\Delta_{\ini\fin}\Delta_{\beta\fin}(1-x)x+\left(m_\beta^2 (\Delta_{\ini \fin}+\Delta_{\beta\fin}) + \Delta_{\fin\beta} (\Delta_{\ini h}+m_\beta^2) x + 
   m_\fin^2 \Delta_{\beta \ini} x^2\right)\log\Lambda_{\ini\fin \beta h }\right], \nonumber\\
   K_{\fin\ini}^{(8)}&=\int dx\frac{m_\ini}{\Delta_{\fin\ini}^2 m_W^2 x} \left[\Delta_{\ini\fin}\Delta_{\beta\fin}(x-1)x-\left(m_\beta^2 \Delta_{\beta\ini}+(m_\fin^2\Delta_{\ini\beta}+m_\beta^2\Delta_{\ini\beta}+m_Z^2\Delta_{\beta\fin})x+m_\fin^2\Delta_{\beta\ini}x^2\right)\log \Lambda_{\ini\fin \beta Z  }\right], \nonumber\\
   K_{\fin\ini}^{(9)}&=\int dx\frac{2m_\ini}{\Delta_{\fin\ini}^2 x}\left[\Delta_{\ini \fin}(x-1)x-\left(m_\beta^2 + (\Delta_{Z\ini}-m_\beta^2)x +m_\fin^2 x^2\right)\log\Lambda_{\ini\fin \beta Z}\right], \nonumber\\
      K_{\fin\ini}^{(10)}&=\int dx\,\frac{4 m_\beta}{\Delta_{\fin \ini}} \log\Lambda_{\ini\fin \beta Z }.
\end{align}

Now that we have the analytical results for the right-- and left--handed neutral boson diagram form factors shown in Eq.~\eqref{eq:form_factors}, we may sum these with the chiral components of ${f}^{(c.c)}_{\fin \ini, \alpha}$ and ${f}^{\rm(g)}_{\fin \ini, \alpha}$ shown in Eq.~\eqref{eq:SM_form_factor_relations} for radiative decays with a photon and a gluon in the final state respectively. More explicitly, this can be written
\begin{align}
     {f}_{\fin \ini} &= {f}^{ (l)}_{\fin \ini} +{f}^{(FCN)}_{\fin \ini}, \nonumber \\
     {f}^{(FCN)}_{\fin \ini}&={f}^{(7)}_{\fin \ini}+{f}^{(8)}_{\fin \ini}+{f}^{(9LL)}_{\fin \ini}+{f}^{(9LR)}_{\fin \ini}+{f}^{(9RL)}_{\fin \ini}+{f}^{(9RR)}_{\fin \ini},
\end{align}
where we have collected all the contributions from the diagrams with FCN vertices into one form factor ${f}^{(FCN)}_{\fin \ini}$. Note that we don't denote the the chirality specifically so the above expression applies for all required combinations $f^{\rm{R}}_{ \fin \ini}, {f}^{\rm{L}}_{ \fin \ini}, {f}^{\rm{R}}_{ \ini  \fin}$ and ${f}^{\rm{L}}_{\ini\fin}$ needed to compute the $CP$ asymmetries outlined in the next section. 

\subsection{Relations between \texorpdfstring{$CP$}{TEXT} asymmetries and form factors\label{subsec:CP}} 

We now briefly outline the $CP$ asymmetries between the radiative decay $q_\ini \to q_\fin+ \gamma_+$ and its $CP$ conjugate process $\bar{q}_\ini \to \bar{q}_\fin+ \gamma_-$.\footnote{It should be noted that the following sections apply for both photons and gluons in the final state, but we present formulae for the photon case for simplicity.} As shown in Ref.~\cite{Balaji:2019fxd}, we get
\begin{align}
\Delta_{CP,+} &= \frac{\Gamma(q_\ini \to q_\fin+ \gamma_+) - 
\Gamma(\bar{q}_\ini \to \bar{q}_\fin+ \gamma_-)}
{\Gamma(q_\ini \to q_\fin+ \gamma) + 
\Gamma(\bar{q}_\ini \to \bar{q}_\fin+ \gamma)} \,, &\Delta_{CP,-} &= \frac{\Gamma(q_\ini \to q_\fin+ \gamma_{-})
- \Gamma(\bar{q}_\ini \to \bar{q}_\fin+ \gamma_{+})}
{\Gamma(q_\ini \to q_\fin+ \gamma) + 
\Gamma(\bar{q}_\ini \to \bar{q}_\fin+ \gamma)}.
\end{align}
The $CP$ asymmetry between $q_\ini \to q_\fin+ \gamma_-$ and its $CP$ conjugate process $\bar{q}_\ini \to \bar{q}_\fin+ \gamma_+$, $\Delta_{CP,-}$, is defined by exchanging $+$ and $-$ signs.
The photon polarisation independent $CP$ asymmetry is obtained by summing $\Delta_{CP,+}$ and $\Delta_{CP,-}$ together which yields
\begin{eqnarray}
\Delta_{CP} &=& \frac{\Gamma(q_\ini \to q_\fin+ \gamma_{+}) - \Gamma(\bar{q}_\ini \to \bar{q}_\fin+ \gamma_{-}) + \Gamma(q_\ini \to q_\fin+ \gamma_{-})
- \Gamma(\bar{q}_\ini \to \bar{q}_\fin+ \gamma_{+}) }
{\Gamma(q_\ini \to q_\fin+ \gamma) + 
\Gamma(\bar{q}_\ini \to \bar{q}_\fin+ \gamma)} \,.
\end{eqnarray}
Since the phase spaces are the same for quark and antiquark channels, these formulae can be written
\begin{align} \label{eq:CP_v1}
\Delta_{CP,+} &= \frac{|f^{\text{L}}_{\fin \ini }|^2 - |f^{\text{R}}_{\ini \fin}|^2}
{|f^{\text{L}}_{\fin \ini }|^2 + |f^{\text{R}}_{\fin \ini }|^2 + |f^{\text{R}}_{\ini \fin}|^2 + |f^{\text{L}}_{\ini \fin}|^2} \,, &\Delta_{CP,-} &= \frac{|f^{\text{R}}_{\fin \ini }|^2 - |f^{\text{L}}_{\ini \fin}|^2}{|f^{\text{L}}_{\fin \ini }|^2 + |f^{\text{R}}_{\fin \ini }|^2 + |f^{\text{R}}_{\ini \fin}|^2 + |f^{\text{L}}_{\ini \fin}|^2} \,,
\end{align}
as well as the sum
\begin{eqnarray} \label{eq:CP_v2}
\Delta_{CP} &=& \frac{|f^{\text{L}}_{\fin \ini }|^2 + |f^{\text{R}}_{\fin \ini }|^2 - |f^{\text{R}}_{\ini \fin}|^2 - |f^{\text{L}}_{\ini \fin}|^2}
{|f^{\text{L}}_{\fin \ini }|^2 + |f^{\text{R}}_{\fin \ini }|^2 + |f^{\text{R}}_{\ini \fin}|^2 + |f^{\text{L}}_{\ini \fin}|^2} \,. 
\end{eqnarray}
We will use the formulation of $\Delta_{CP,+}$ and $\Delta_{CP,-} $ in the phenomenological studies that follow.

\section{Results}
\label{sec:results}
Since the Lorentz invariant amplitude can be written
\begin{align}
     i\mathcal{M} (q_\ini \to q_\fin + \gamma)  = i \bar{u}(p_\beta) \sigma^{\mu \nu}(A_\gamma+B_\gamma \gamma_5)q_\nu u(p_t) \varepsilon^*_{\pm,\mu}(q) \,,
\end{align}
we may express the decay width in terms of the vector and axial form factors $A$ and $B$
\begin{align}
    \Gamma(q_\ini\rightarrow q_\fin\gamma)&=\frac{1}{\pi}\left(\frac{m_\ini^2-m_\fin^2}{2m_\ini}\right)^3\left(|A_\gamma|^2+|B_\gamma|^2\right),&\Gamma(q_\ini\rightarrow q_\fin g)&=\frac{C_F}{\pi}\left(\frac{m_\ini^2-m_\fin^2}{2m_\ini}\right)^3\left(|A_g|^2+|B_g|^2\right),
\end{align}
where $\gamma$ and $g$ simply denotes processes with a photon or gluon in the final state and $C_F=4/3$ is the standard colour factor. Comparing the vector and axial form factors with their chiral counterparts shown in Eq.~\eqref{eq:SM_form_factor_relations}, we get the simple relations
\begin{align}
     A &= \frac{f^{\text{L}}_{\fin \ini }+f^{\text{R}}_{\fin \ini }}{2}, &B&=\frac{f^{\text{R}}_{\fin \ini }-f^{\text{L}}_{\fin \ini }}{2}.\nonumber
\end{align}
We may then compute the branching ratio ${\cal B}(q_\ini\to q_\fin \gamma) = \Gamma(q_\ini\to q_\fin \gamma)/ \Gamma_{q_\ini}$ for the radiative process. For the total decay width $\Gamma_{q_\ini}$, we include the three dominant tree--level decay channels. In the case of an initial state $t'(b')$, these would be $t'(b')\to d_i(u_i) W^+$, $u_i(d_i)Z$ and $u_i(d_i)h$ where  $u_i$ and $d_i$ are any up-- or down--type quarks that are kinematically allowed in the decays \cite{Cacciapaglia:2010vn}. 

In general, we expect that the branching ratios scale like $|X_{\fin\ini}|^2$ and $|\Delta_{CP,+}|\ll |\Delta_{CP,-}|$, because of helicity suppression which arises due to angular momentum conservation and the fact that the charged current interaction is parity violating \cite{Balaji:2020qjg}. The exact scaling of the $CP$ asymmetries with the addition of the VLQ to the SM, is non-trivial and highly model dependent due to fine interplay between quark masses and the quartet of quark mixing angles that enters in the squared Lorentz invariant amplitudes. 


\subsection{\texorpdfstring{$CP$}{TEXT} violation with vector isosinglet up--type quark }
\label{sec:CPuptype}
\subsubsection{New physics signals with vector isosinglet \texorpdfstring{$t'$}{TEXT}}
\label{sec:newsignalsuptype}
We may now study the phenonemological implications of a $t'$ quark by analysing a global fit that combines a variety of complementary flavour physics observables to constrain the quark mixing matrix  \cite{Alok:2015iha}. These global fits show that the extended quark mixing matrix is very strongly constrained, but there can be enhancements to flavour changing decays that would otherwise be extremely suppressed in the SM. In the fit considered in Ref.~\cite{Alok:2015iha}, the new physics effects of the $t'$ are mainly through charged current interactions which involve quark mixing via the matrix $V$ and neutral current and Higgs interactions that mix through the matrix $X^u=VV^\dagger$. The extended CKM matrix $V$ in such a setup comprises four SM and five new physics parameters. Flavour physics data is used to perform a combined fit to these parameters and the best fit values of the parameters indicates whether the new physics parameters can be nonzero.

For the fit, in addition to the six directly measured CKM matrix elements, additional flavour physics observables with small hadronic uncertainties are included, they then consider two benchmark masses $m_{t'}=800$ GeV and $m_{t'}=1200$ GeV \cite{Alok:2015iha}.

\begin{table}[!ht]
\centering
  \begin{tabular}{ |c | c | c | c | }
    \hline
    Parameter & SM &$m_{t'}=800$ GeV&$m_{t'}=1200$ GeV \\ \hline
    $\lambda$ & $0.226\pm 0.001$ &  $0.226\pm 0.001$ & $0.226\pm 0.001$\\ 
    $A$ &  $0.780\pm 0.015$ &  $0.770\pm 0.019$&$0.769\pm 0.019$ \\
       $C$ & $0.39\pm 0.01$ & $0.44\pm 0.02$&$0.43\pm 0.02$ \\
       $\delta_{13}$&$1.21\pm0.08$&$1.13\pm0.11$&$1.15\pm0.09$\\ \hline
       $P$& - &$0.40 \pm 0.26 $ &$0.30 \pm 0.21$\\
       $Q$& - & $0.04 \pm 0.06 $&$0.03 \pm 0.05$\\
       $R$& - &$0.45 \pm 0.25$ &$0.36 \pm 0.22$\\
       $\delta_{41}$&-& $0.55 \pm 0.45$&$0.76 \pm 0.42$\\
       $\delta_{42}$&-&$0.52 \pm 3.26 $ &$0.96 \pm 1.21$\\\hline
       $\chi^2$/ d.o.f.&$71.15/60 $&$63.35/59 $&$63.60/59$\\
    \hline
  \end{tabular}
          \caption{Best fit parameters for the quark mixing matrix with a vector--like quark isosinglet $t'$ included. These are shown for two benchmark quark masses $m_{t'}=800~\rm GeV$ and $m_{t'}=1200~\rm GeV$ outlined in Table 4 of Ref.~\cite{Alok:2015iha}.}
    \label{tab:wolfenstein_bestfit}
\end{table}

The quark mixing matrix $V$ corresponds to the $4\times 3$ sub matrix of Eq.~\eqref{eq:CKM4}. However it is convenient to use the Hou--Soni--Steger parametrisation as shown in Appendix~\ref{sec:Wolfenstein} which contains four SM parameters $\lambda$, $A$, $C$, $\delta_{13}$ and introduces five new physics parameters $P$, $Q$, $r$, $\delta_{41}$, $\delta_{42}$ \cite{Hou:1987hm}. The fit considers all observables that can constrain the parameters of the quark mixing matrix. The total $\chi^2$ is evaluated as a function of these parameters, and the subsequent best fit values are chosen to minimise this function. The resulting Wolfenstein parameters and resulting $\chi^2$ goodness--of--fit are given in Table~\ref{tab:wolfenstein_bestfit}.

Now that we have the best fit values for the mixing matrix, we may apply the formulae developed in this work to compute the radiative decays of the $t'$ or $t$ quarks into light up--type quarks in conjunction with a gluon or photon. We are particularly interested in the branching ratio $\mathcal{B}$ and the $CP$ ratios $\Delta_{CP,+}$ and $\Delta_{CP,-}$ as defined in Section~\ref{subsec:CP}. The results for these three observables corresponding to each kinematically allowed decay are shown in Table~\ref{tab:vlqu_results}. 

One important consideration we have included is a proper computation of the quark masses in the loops and also the strong coupling constant (in the case of gluon emission) at an appropriate renormalisation scale using the minimal subtraction $\overline{\textrm{MS}}$ scheme. The physical boson masses are fixed at $m_Z=91.19$ GeV, $m_W=80.39$ GeV and $m_h=125.1$ GeV \cite{Zyla:2020zbs}, while the quark masses are run to the scale of the decaying particle at two--loop order with five active flavours $n_f=5$ using the \texttt{RunDec} package \cite{Chetyrkin:2000yt,Herren:2017osy}. For example, if we consider the effect of the mass evolution of the ``heavier" second and third generation quarks, where we use the running $\overline{\textrm{MS}}$ masses  $(m_c(m_c),m_b(m_b))=(1.27,4.18)$ GeV \cite{Zyla:2020zbs}, where the reference values are at a scale denoted by the masses themselves while the strange quark mass at a scale of $2$ GeV is given by $m_s(\mu\approx 2)=0.095$ GeV. The top quark pole mass is set to $M_t=173.21$ GeV, then for example, for a $t'$ decaying with mass $m_{t'}=800$ GeV, the heavy quarks in the various loops are assigned masses $m_b(\mu=m_{t'})=2.460$ GeV, $(m_c(\mu=m_{t'}),m_t(\mu=m_{t'}))=(0.595, 155.76)$ GeV. The QCD coupling is given by $\alpha_s(m_Z)=0.1181$ \cite{Zyla:2020zbs} so using the same prescription as for the quark masses, at $m_{t'}=800$ GeV we get $\alpha_s(\mu=m_{t'})=0.089$. The quark masses and strong coupling constant at the decaying quark mass scale is used rather than at pole masses for calculations in this section and all those following.



We find that the largest branching ratio for radiative decays of $t'$ for both $m_{t'}=800$ GeV and $m_{t'}=1200$ GeV is for the $t'\rightarrow tg$ process which has a relatively large $\mathcal{B}\simeq10^{-5}$ while decays to $ug$ and $cg$ final states are similar with $\mathcal{B}\simeq 10^{-8}\text{-}10^{-7}$. We note that the ratio $\frac{\mathcal{B}(t'\rightarrow c\gamma(g)} {\mathcal{B}(t'\rightarrow u\gamma(g) )}\simeq 1$ based on the parameters in Table~\ref{tab:wolfenstein_bestfit}. Decays to photon final states are about one order of magnitude smaller due to the weakness of the electromagnetic coupling constant. However the decay $t'\rightarrow t\gamma$ is phenomenological relevant due to its relatively clean experimental signature with a photon in the final state. We find the subdominant $CP$ ratio $|\Delta_{CP,+}|$ having absolute order of magnitudes for the central values chosen in the fit for $t'$ decays into $g$ with $|\Delta_{CP,+}|\simeq 10^{-8}\text{-}10^{-7}, 10^{-8}, 10^{-14}$ with $t$, $c$ and $u$--quarks in the final state respectively. While the larger $CP$ ratio has absolute magnitudes at the order of $|\Delta_{CP,-}|\simeq 10^{-8}, 10^{-4}, 10^{-4}$ for $t$, $c$ and $u$--quark emitted in conjunction with a gluon. The fact that up to $\simeq10^{-4}$ $CP$ asymmetries can be generated is striking. Similarly for $t'$ decays into a photon we get $|\Delta_{CP,+}|\simeq 10^{-8}, 10^{-9}, 10^{-15}$ and $|\Delta_{CP,-}|\simeq 10^{-7}, 10^{-5}, 10^{-6}$ with $t$, $c$ and $u$--quarks in the final state respectively. We find for photon emission that the largest $CP$ asymmetry is generated with $c$ or $u$--quarks in the final state respectively. 

We note that for these parameters, the top quark radiative decay widths are suppressed, however they are orders of magnitude larger than the SM predictions \cite{Balaji:2020qjg}. The largest branching ratio is for $\mathcal{B}(t\rightarrow ug )\simeq 10^{-11}$ which is still well below experimental reach. However, interestingly we note that the $CP$ ratio across the top quark decay channels can have $|\Delta_{CP,-}|\simeq 10^{-3}\text{-}10^{-1}$ for the two benchmark $t'$ masses and can even achieve unity. We find that $|\Delta_{CP,+}|$ is smaller and lies in the range $|\Delta_{CP,+}|\simeq 10^{-13}\text{-}10^{-4}$ when considering all the channels. 

\begin{table}[!h]
    \centering
    \begin{tabular}{|c|c|c|c|} 
    \hline
    \multirow{2}{15mm}{Decay channel} &  \multicolumn{3}{c|}{$m_{t'}=800~{\rm GeV}$} \\\cline{2-4}
    &   $\mathcal{B}\qquad$ & $\Delta_{CP,+}\qquad$ &  $\Delta_{CP,-}\qquad$ \\\hline
     $t'\to u \gamma$& $ (1.2^{+7.0}_{-1.1}\times10^{-8}$ &  $(-0.6^{+2.3}_{-4.0})\times10^{-14}$ &$(-0.5^{+2.5}_{-6.0})\times10^{-5}$  \\\hline
     $t'\to c \gamma$& $(1.0^{+2.4}_{-1.0})\times10^{-8}$ &$(0.1^{+3.5}_{-8.0})\times10^{-8}$ & $0.02^{+1.40}_{-2.63}\times10^{-3}$ \\\hline
     $t'\to t \gamma$&$ (3.36^{+0.03}_{-0.06})\times10^{-6} $ &$(-0.04^{+2.09}_{-1.00})\times10^{-6} $ &$(-0.09^{+2.9}_{-1.9}) \times10^{-6}$ \\\hline
     $t'\to u g$&$ (0.5^{+2.0}_{-0.4} )\times10^{-7} $ &$(0.7^{+7.0}_{-1.3})\times10^{-13}$ &$(1.1^{+9.0}_{-1.8})\times10^{-4}$  \\\hline
     $t'\to c g$&$ (0.4^{+2.1}_{-0.4})\times10^{-7} $ &$(-0.01^{+0.2}_{-1})\times10^{-6}$ &$(  -0.02^{+0.40}_{-1.82})\times10^{-2}$\\\hline
     $t'\to t g$&$ 3.937^{+0.077}_{-0.109}\times10^{-5} $ &$(-0.09^{+4.05}_{-2.90})\times10^{-6}$ &$(0.11^{+8.07}_{-6.00})\times10^{-7}$  \\\hline
     $t\to u \gamma$&$(0.29^{+1.1}_{-0.28})\times10^{-11} $ &$(0.7^{+4.0}_{-0.7})\times10^{-12}$ &$(-0.60^{+0.70}_{-2.80})\times10^{-2}$  \\\hline
     $t\to c \gamma$&$(0.7^{+5.0}_{-0.7})\times 10^{-12} $ & $(-0.04^{+0.8}_{-1.1})\times10^{-5} $ &$0.04^{+0.90}_{-0.92} $  \\\hline
     $t\to u g$&$ (0.7^{+2.4}_{-0.6})\times10^{-10}$ &$(1.4^{+8.0}_{-1.5})\times10^{-12}$ &$-0.012^{+0.011}_{-0.070}$  \\\hline
     $t\to c g$&$ (0.13^{+2.00}_{-0.11})\times10^{-10} $ &$(-0.1^{+1.3}_{-0.9})\times10^{-5}$ &$0.1^{+0.7}_{-1.1}$ \\\hline
    \end{tabular}
    \end{table}
    \begin{table}[!h]
    \centering
        \begin{tabular}{|c|c|c|c|} 
     \hline
     \multirow{2}{15mm}{Decay channel} &  \multicolumn{3}{c|}{$m_{t'}=1200~{\rm GeV}$} \\\cline{2-4}
    &   $\mathcal{B}\qquad$ & $\Delta_{CP,+}\qquad$ &  $\Delta_{CP,-}\qquad$ \\\hline
     $t'\to u \gamma$& $ (0.8^{+0.6}_{-0.7})\times 10^{-8}$ &$(-0.22^{+0.27}_{-1.6}\times10^{-14})$ & $(0.1^{+2.0}_{-2.3})\times10^{-5}$  \\\hline
     $t'\to c \gamma$& $ (0.6^{+5.0}_{-0.5})\times10^{-8}$ &$(0.7^{+9.0}_{-5.0})\times10^{-9}$ &$(0.3^{+3.0}_{-2.7})\times10^{-4}$ \\\hline
     $t'\to t \gamma$&$(2.71^{+0.01}_{-0.06})\times10^{-6} $ &$(-0.8^{+4.0}_{-8.0})\times10^{-7}$ &$ (-0.2^{+1.2}_{-1.4})\times 10^{-6}$ \\\hline
     $t'\to u g$&$(0.26^{+1.7}_{-0.19})\times10^{-7}$ &$(0.3^{+2.2}_{-0.4})\times10^{-13}$ &$(0.5^{+2.3}_{-0.7})\times10^{-4}$ \\\hline
     $t'\to c g$&$(0.21^{+1.1}_{-0.14})\times10^{-7}$ &$(-0.1^{+0.8}_{-1.3})\times10^{-7}$ &$(-0.3^{+2.4}_{-2.3})\times10^{-3}$ \\\hline
     $t'\to t g$&$(4.21^{+0.54}_{-1.08})\times10^{-5}$ &$(-1.0^{+7.0}_{-8.0})\times10^{-7}$ &$(0.2^{+4.0}_{-2.3})\times10^{-7}$ \\\hline
     $t\to u \gamma$&$(1.0^{+4.0}_{-1.0})\times10^{-12}$ &$(0.8^{+2.7}_{-1.2})\times10^{-12}$ &$-0.006^{+0.011}_{-0.018}$ \\\hline
     $t\to c \gamma$&$(0.22^{+2.7}_{-0.22})\times10^{-12}$ &$(-0.2^{+1.2}_{-0.7})\times10^{-5}$ &$0.2^{+0.6}_{-1.0}$ \\\hline
     $t\to u g$&$(0.24^{+1.30}_{-0.21})\times10^{-10}$ &$(0.15^{+1.3}_{-0.34})\times10^{-11}$ &$-0.014^{+0.03}_{-0.11}$ \\\hline
     $t\to c g$&$(0.5^{+6.0}_{-0.4})\times10^{-11}$ &$(-0.6^{+1.8}_{-0.4})\times10^{-5}$ &$0.61^{+0.28}_{-1.70}$\\ \hline
    \end{tabular}
    \caption{Results for the branching ratios and $CP$ asymmetries for the  radiative decay channels of the vector--like quark $t'$ and top quark. These are based on the best fit mixing parameters and benchmark masses $m_{t'}=800~\rm GeV$ (top) and $m_{t'}=1200~\rm GeV$ (bottom) outlined in Table \ref{tab:wolfenstein_bestfit}.}
    \label{tab:vlqu_results}
\end{table}
\subsubsection{\texorpdfstring{$CP$}{TEXT} observables for vector \texorpdfstring{$t'$}{TEXT} addressing the CKM unitarity problem}
\label{sec:CKMunitarityproblem}
\paragraph{Radiative decay observables at the low--$\chi^2$ benchmark}\mbox{}\\\\
It has been pointed out that hints of deviations from unitarity in the first row of the CKM matrix may be explained by the presence of a single vector--like top \cite{Belfatto:2021jhf,Branco:2021vhs}. This solution is particularly attractive since the experimental limits from FCN currents in the up quark sector are less stringent than those in the down sector. The fit from the study considered in this section explored how the stringent experimental constraints arising from $CP$ Violation in the kaon sector and from meson mixing such as $D^0{\text -}\overline{D}^0$, $K^0{\text -}\overline{K}^0$ and $B_{ds}^0{\text -}\overline{D}_{ds}^0$ could be simultaneously satisfied. It ultimately concluded that in order to concurrently explain the deviations from unitarity while maintaining perturbativity of Yukawa couplings, that $t'$ should have a mass $m_{t'}<7$ TeV. 

This is noteworthy, since the existence of such states, if sufficiently light, are within reach of upcoming experiments in the high energy frontier including the LHC and its upgrade. The largest contributions to the pair production cross sections only depend on the mass $m_{t'}$ \cite{DeSimone:2012fs}. However, single $t'$ production mechanisms can in part be approximated as functions of the mixing matrix elements $V_{t'd}$, $V_{t's}$ and $V_{t'b}$ and are therefore more model dependent \cite{Buchkremer_2013}. These same mixing parameters control the rate of the $t'$ decay into vector bosons and SM quarks \cite{Botella_2017}. Lower bounds on the VLQ mass, $m_{t'}>1.3$ TeV \cite{Aaboud_2018} and $m_{t'} \gtrsim 1.0$ TeV \cite{Sirunyan_2019}, have been obtained at the $95\%$ CL by the ATLAS and CMS detectors, respectively, in searches for pair produced $t'$ quarks. However it should be noted that these searches assume the new VLQ only couples to the third SM generation and therefore require reinterpretation when mixing with all generations is considered.

The low $\chi^2=3.2$ benchmark in the parameter scan performed corresponds to Eq. (4.3) in Ref.~\cite{Branco:2021vhs}, which we repeat here for convenience,
\begin{align}
\label{eq:vlqtplowchi}
    \theta_{12} &= 0.2265, &\theta_{13} &= 0.003818, &\theta_{23} &= 0.03988,  \nonumber\\
    \theta_{14} &= 0.03951, &\theta_{24} &= 0.002078, &\theta_{34} &= 0.01271,  \nonumber\\
    \delta_{13} &=0.0396\pi, &\delta_{14} &= 1.818\pi, &\delta_{24} &= 0.728\pi,
\end{align}
with $m_{t'}=1.5$ TeV and a perturbativity factor of $p-1\simeq 0.13$. The $2\sigma$ fit region has $|V_{41}|>|V_{42}|,|V_{43}|$, meaning that $t'$ couples more strongly with the first generation than the second or third, challenging standard assumptions.  We substitute \eqref{eq:vlqtplowchi} into the general $4\times4$ quark mixing matrix shown in Appendix \ref{sec:quarkmixingmatrix} and then take the $4\times 3$ submatrix to parameterise the quark mixing with the $t'$. Here, we are interested in studying the $CP$ observables for the radiative decays of $t'$ and $t$ quarks and analyse how they change as a function of $m_{t'}$. However, it should be noted that the mixing parameters will be fixed as shown in Eq.~\eqref{eq:vlqtplowchi} for simplicity. We find that for the selected point we recover the tree--level flavour conserving decay $\mathcal{B}(t\rightarrow uZ)\simeq 1.16\times 10^{-7}$ which provides a useful consistency check with Ref.~\cite{Branco:2021vhs}.

\begin{figure}[!h]
    \centering
     \includegraphics[width=0.44\textwidth]{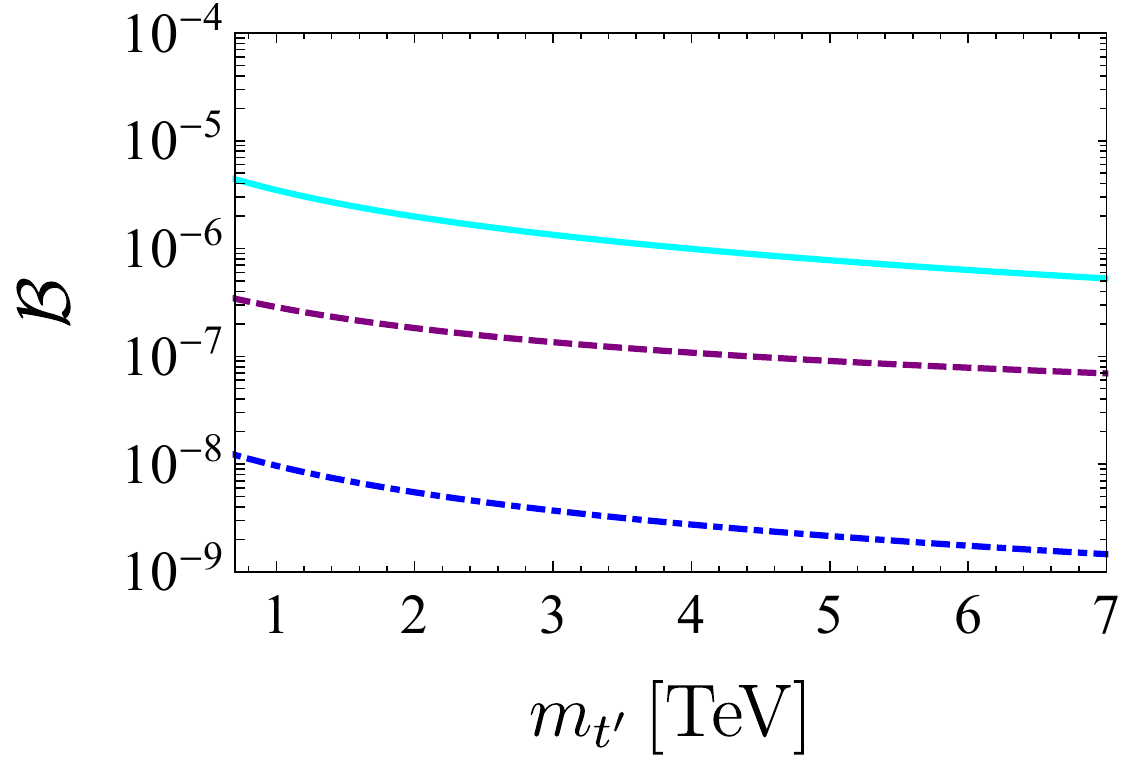}
         \includegraphics[width=0.44\textwidth]{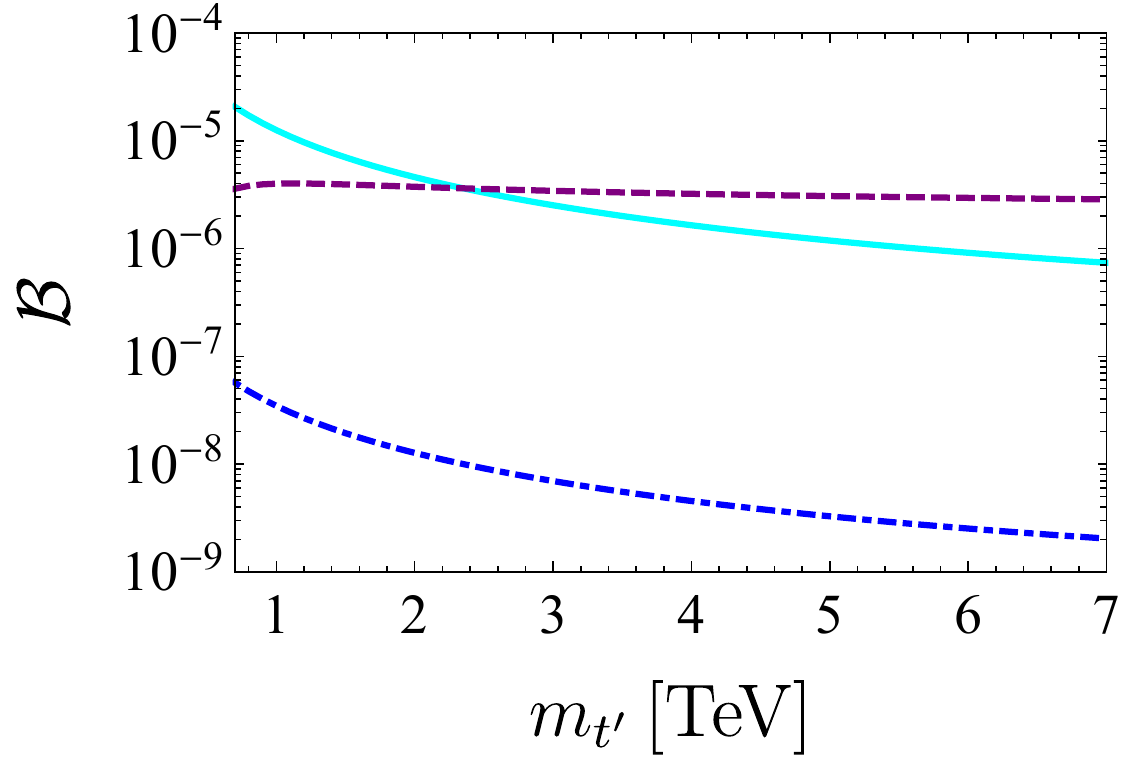}\\
     \includegraphics[width=0.44\textwidth]{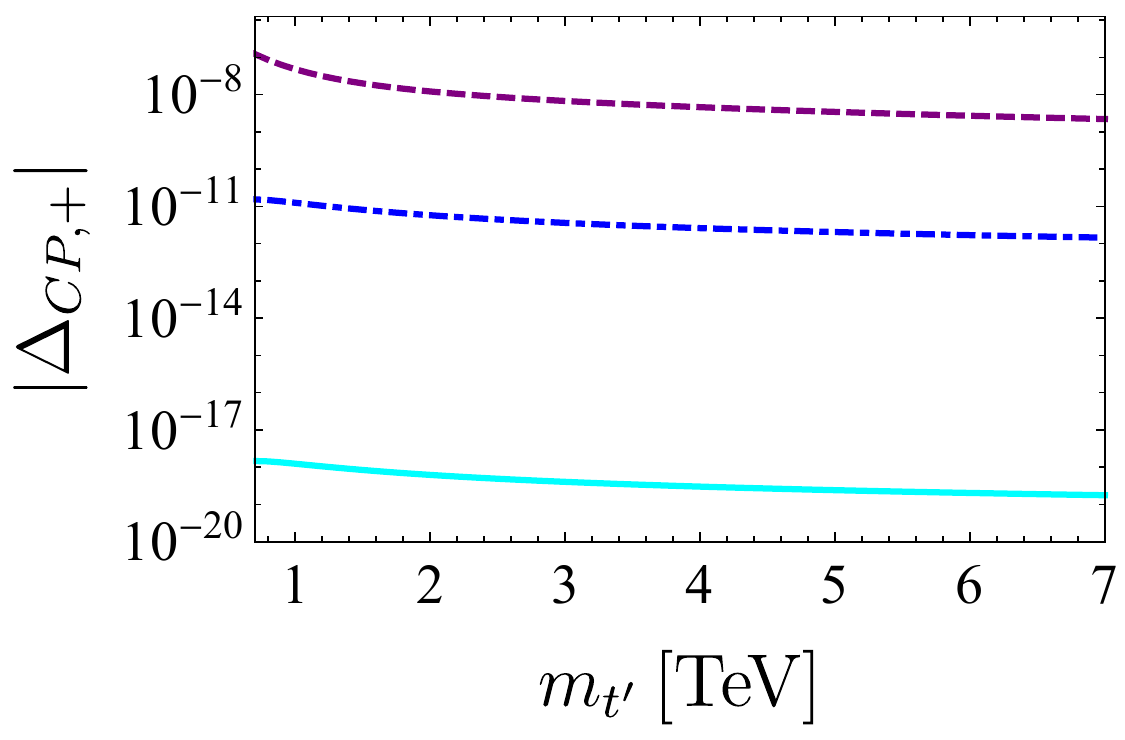}
         \includegraphics[width=0.44\textwidth]{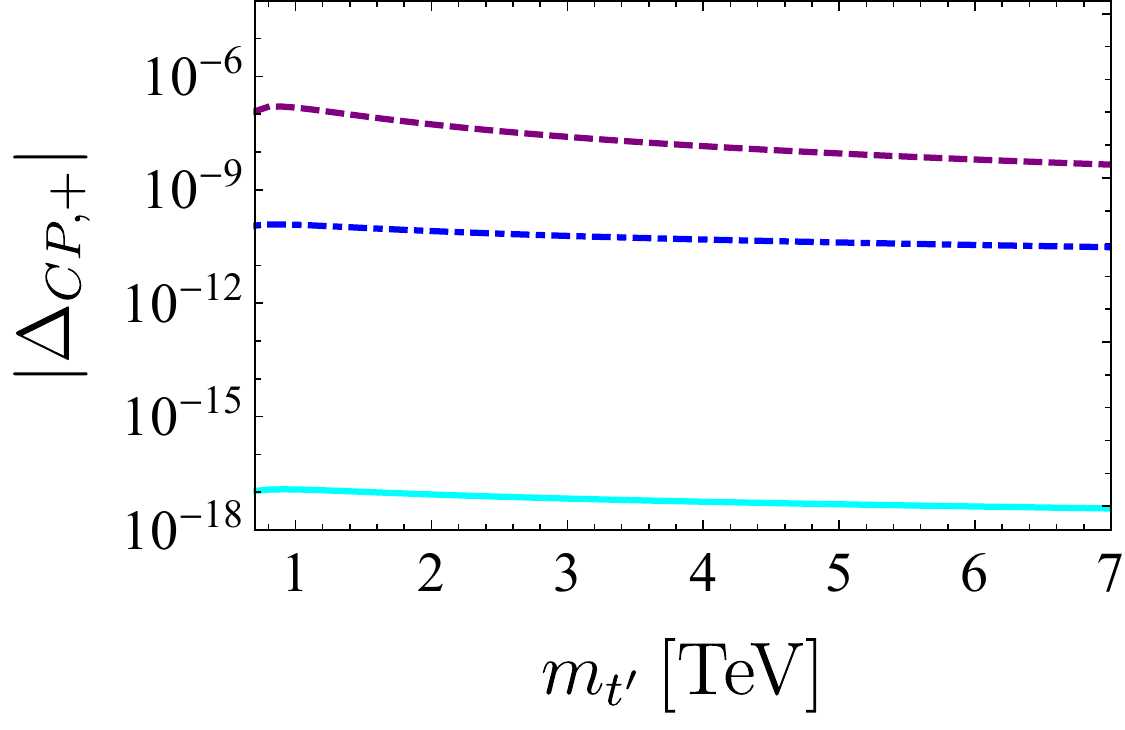}
         \includegraphics[width=0.44\textwidth]{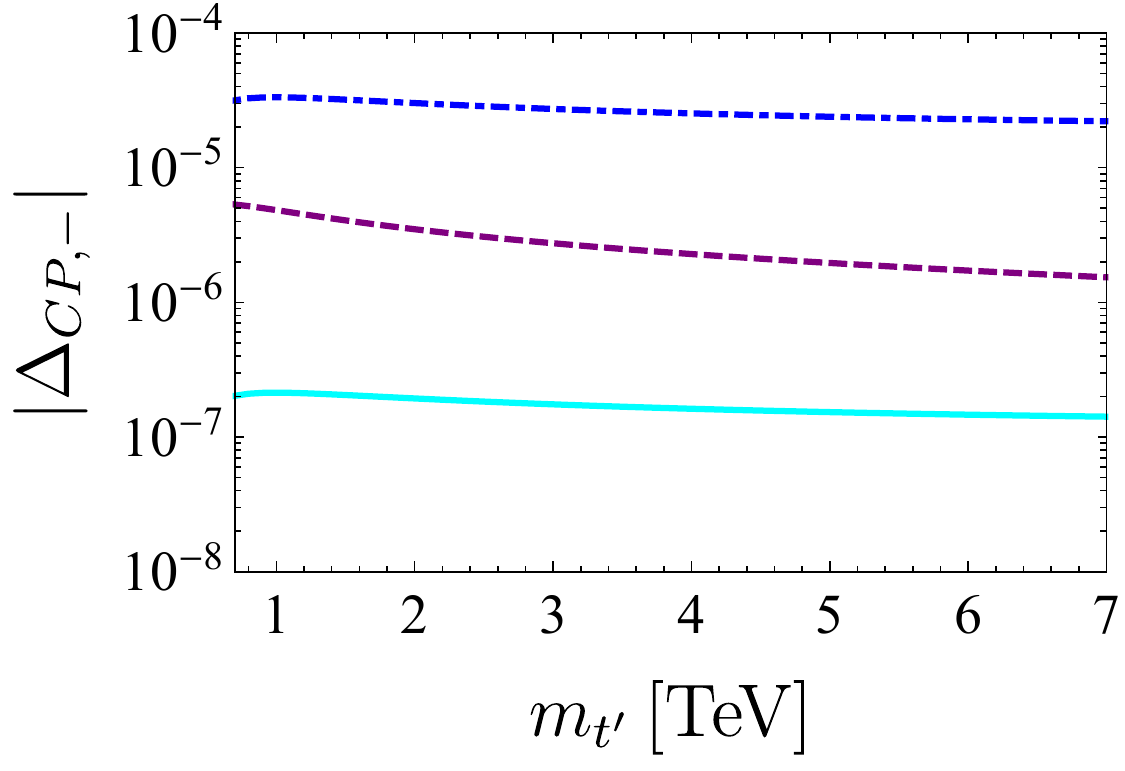}
         \includegraphics[width=0.44\textwidth]{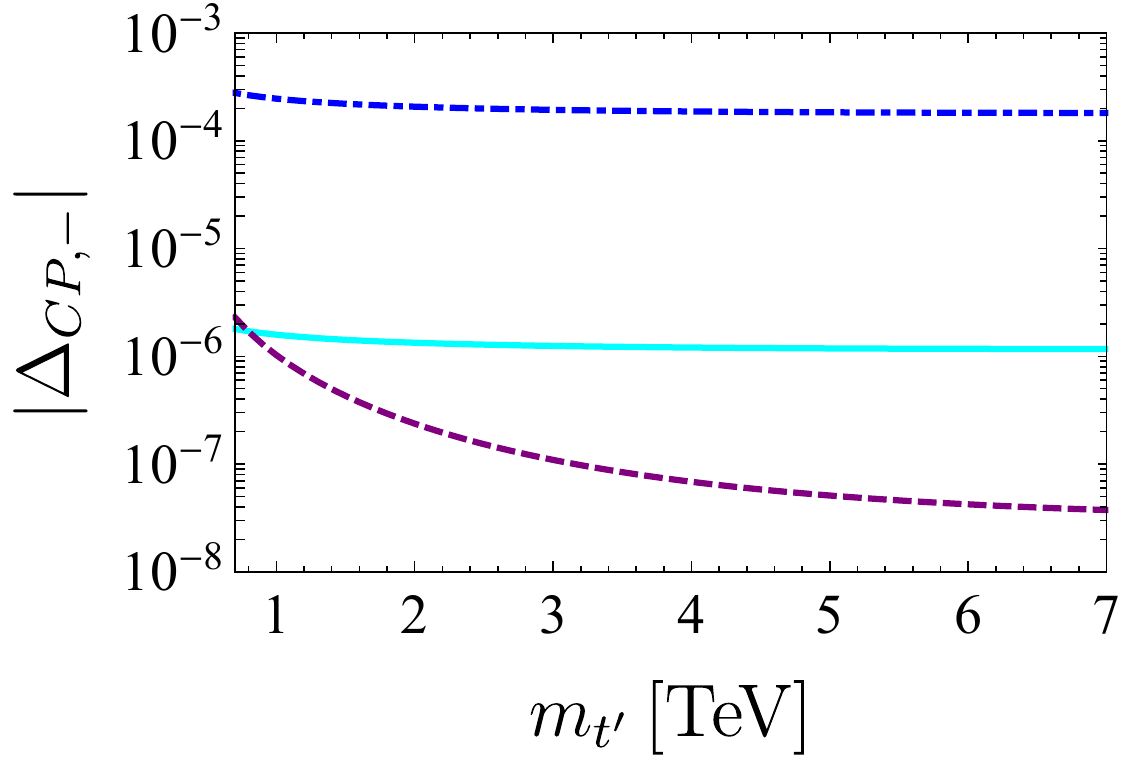}
    \caption{The branching ratio (top panels), the $|\Delta_{CP,+}|$ asymmetry (middle panels) and $|\Delta_{CP,-}|$ asymmetry (bottom panels) for the radiative decays of $t'$. The left (right) panels are for radiative decays with a final state photon (gluon). The solid, dot--dashed and dashed lines correspond to $t'$ decay channels with a $u$, $c$ and $t$ quark final state respectively shown as a function of the vector--like quark mass $m_{t'}$. The low $\chi^2$ benchmark mixing parameters shown in Eq.~\eqref{eq:vlqtplowchi} are used while $m_{t'}=[0.7,7]$ TeV.  }
    \label{fig:brcp}
\end{figure}

The evolution of the branching ratios and $CP$ asymmetries for the $t'\rightarrow u_i\gamma(g)$ channels are shown in Figure~\ref{fig:brcp}. We note from the top left panel of Figure~\ref{fig:brcp} that the branching ratio hierarchy for $\mathcal{B}(t'\rightarrow u \gamma)>\mathcal{B}(t'\rightarrow t \gamma)>\mathcal{B}(t'\rightarrow c \gamma)$ lies in the range $\simeq 10^{-9}\text{-}10^{-5}$ over the whole $m_{t'}$ range considered. This is unsurprising given the mixing matrix elements being arranged $|V_{41}|,|V_{43}|,|V_{42}|=0.0390,0.0065,0.0126$. Considering the top right panel of Figure~\ref{fig:brcp}, we note that the branching ratios follow a similar hierarchy, however at $m_{t'}\simeq 2.4$ TeV, $\mathcal{B}(t'\rightarrow tg)>\mathcal{B}(t'\rightarrow ug)$. This is because, even though the mixing is largest between the VLQ and the first generation, the flavour changing diagrams with the Higgs boson get enhanced by the large Yukawa couplings of the VLQ which eventually exceeds the contributions coming purely from quark mixing. We note that the branching ratios for decays into a gluon final state are unsurprisingly about one order of magnitude larger than into a photon final state similar to the discussion in the previous section.

The $CP$ ratios displayed in the middle panels of Figure~\ref{fig:brcp} show that $|\Delta_{CP,+}(t'\rightarrow u_i\gamma)|\simeq 10^{-18}\text{-}10^{-7}$ and $|\Delta_{CP,+}(t'\rightarrow u_i g)|\simeq 10^{-18}\text{-}10^{-6}$  for $u_i=u,c,t$ quarks in the final state respectively. The largest magnitudes being for both the photon and gluon channels being generated by the $t'\rightarrow t\gamma$ and $t'\rightarrow t g$ processes respectively. Finally, referring to the $CP$ ratios shown in the bottom panels of Figure~\ref{fig:brcp}, we observe  $|\Delta_{CP,-}(t'\rightarrow u_i\gamma)|\simeq 10^{-7}\text{-}10^{-4}$ and $|\Delta_{CP,-}(t'\rightarrow u_i g)|\simeq 10^{-8}\text{-}10^{-3}$  for $u_i=u,c,t$ quarks in the final state respectively. The largest values being for both the photon and gluon channels being generated by the $t'\rightarrow c\gamma$ and $t'\rightarrow c g$ processes respectively. The generation of such large $CP$ asymmetries from these VLQ decays and the simultaneous economical explanation for the CKM unitarity problem provides a compelling case for complementary VLQ searches. It should be noted that Figure~\ref{fig:brcp} only shows the absolute value of the $CP$ asymmetries.\footnote{The signs over the displayed mass range are such that $\Delta_{CP,+}(t'\rightarrow u\gamma),\Delta_{CP,+}(t'\rightarrow t\gamma)>0$ while $\Delta_{CP,+}(t'\rightarrow c\gamma)<0$. $\Delta_{CP,-}(t'\rightarrow u\gamma)<0$ and $\Delta_{CP,-}(t'\rightarrow c\gamma), \Delta_{CP,-}(t'\rightarrow t\gamma)>0$. For the gluon channels we get $\Delta_{CP,+}(t'\rightarrow ug),\Delta_{CP,+}(t'\rightarrow tg)>0$ while $\Delta_{CP,+}(t'\rightarrow cg)<0$ while $\Delta_{CP,-}(t'\rightarrow ug)>0$ and $\Delta_{CP,-}(t'\rightarrow cg), \Delta_{CP,-}(t'\rightarrow tg)<0$.}

The top quark decays with a photon radiated for these fit parameters show minimal $m_{t'}$ dependence with branching ratios and $CP$ asymmetries of
\begin{align}
    &\mathcal{B}(t\rightarrow u\gamma)\simeq 3\times10^{-12}, & &\mathcal{B}(t\rightarrow c\gamma)\simeq 9\times10^{-14}, \nonumber \\
    &|\Delta_{CP,+}(t\rightarrow u\gamma)|\simeq 1.4\times10^{-12}, & &|\Delta_{CP,+}(t\rightarrow c\gamma)|\simeq 1.8\times10^{-6}, \nonumber \\
    &|\Delta_{CP,-}(t\rightarrow u\gamma)|\simeq 9.3\times10^{-3}, & &|\Delta_{CP,-}(t\rightarrow c\gamma)|\simeq 0.14,
\end{align}
across the whole mass range considered. We see that the branching ratio for decay into a $u\gamma$ is larger than a $c\gamma$ final state, but the $CP$ ratios are conversely larger for $c\gamma$ decay products. For a gluon in the final state, we again see little $m_{t'}$ dependence, where
\begin{align}
    &\mathcal{B}(t\rightarrow ug)\simeq 6.4\times10^{-11}, & &\mathcal{B}(t\rightarrow cg)\simeq 6.7\times10^{-12}, \nonumber \\
    &|\Delta_{CP,+}(t\rightarrow ug)|\simeq 2.4\times10^{-12}, & &|\Delta_{CP,+}(t\rightarrow cg)|\simeq 9\times10^{-7}, \nonumber \\
    &|\Delta_{CP,-}(t\rightarrow ug)|\simeq 0.024, & &|\Delta_{CP,-}(t\rightarrow cg)|\simeq 0.11,
\end{align}
which reflects a similar hierarchy when arranged in terms of final state quark flavour to the photon case.In the following section, we consider the allowable region after doing a scan of the full parameter space, including the quark mixing.

\begin{figure}[!h]
    \centering
     \includegraphics[width=0.49\textwidth]{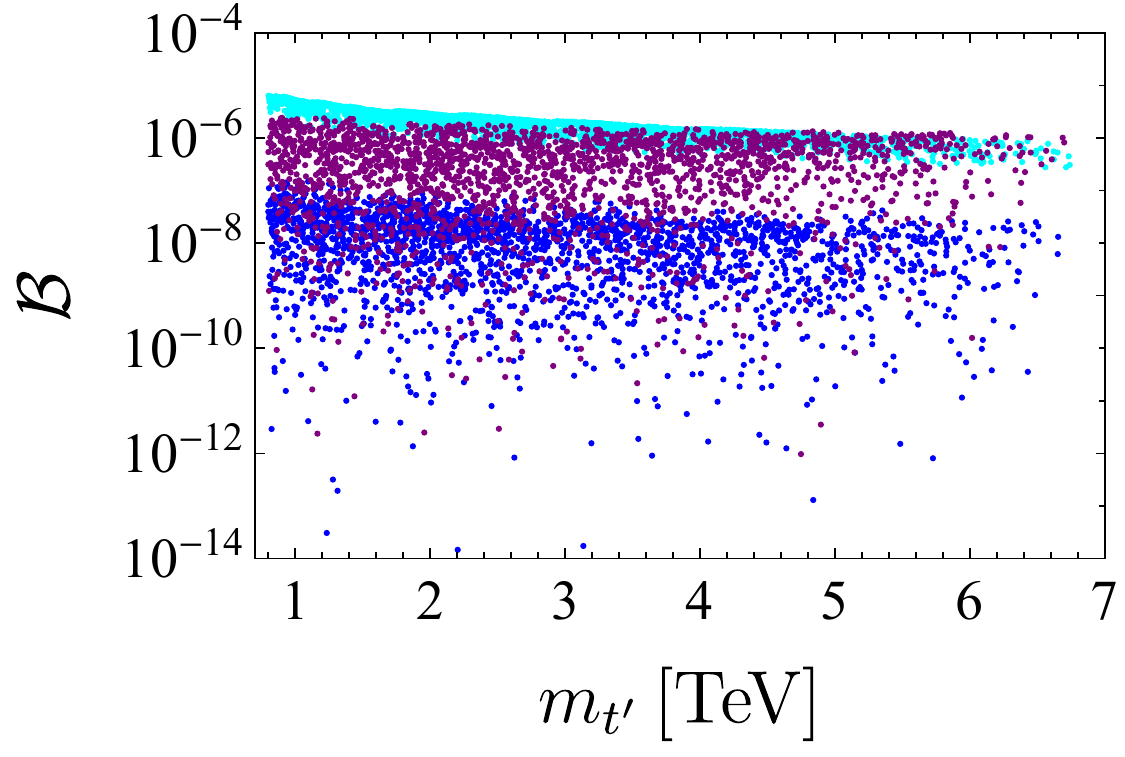}
         \includegraphics[width=0.49\textwidth]{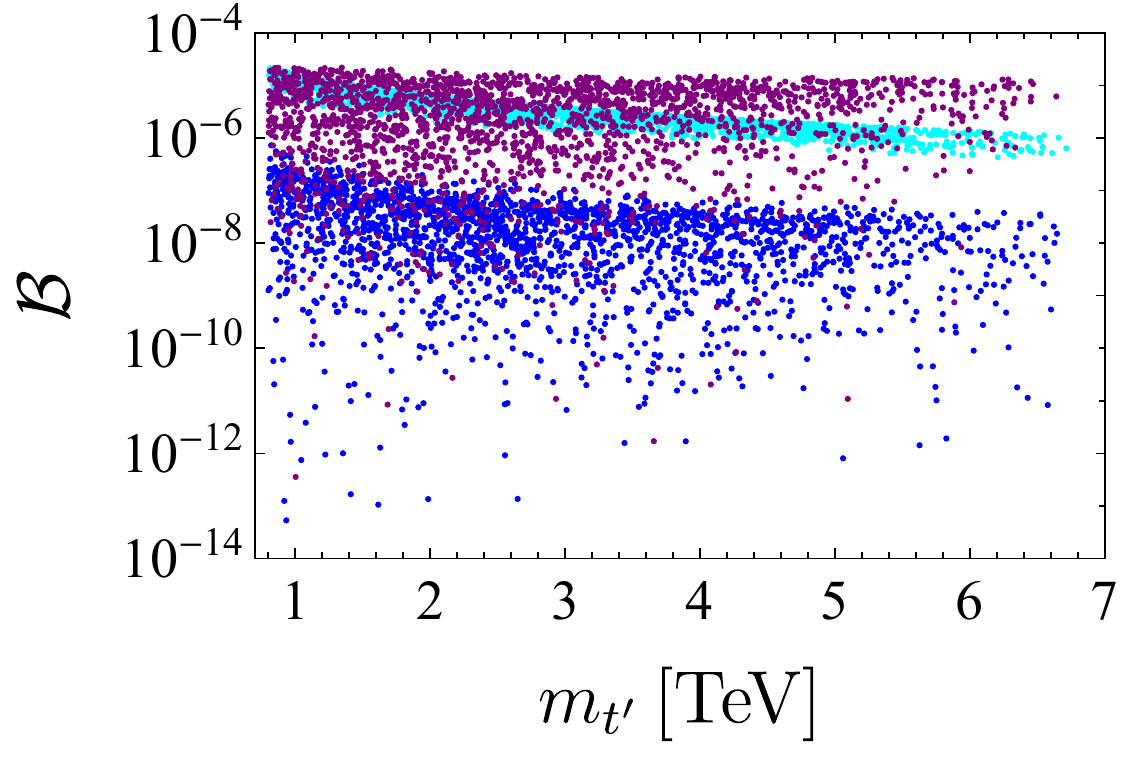}\\
     \includegraphics[width=0.49\textwidth]{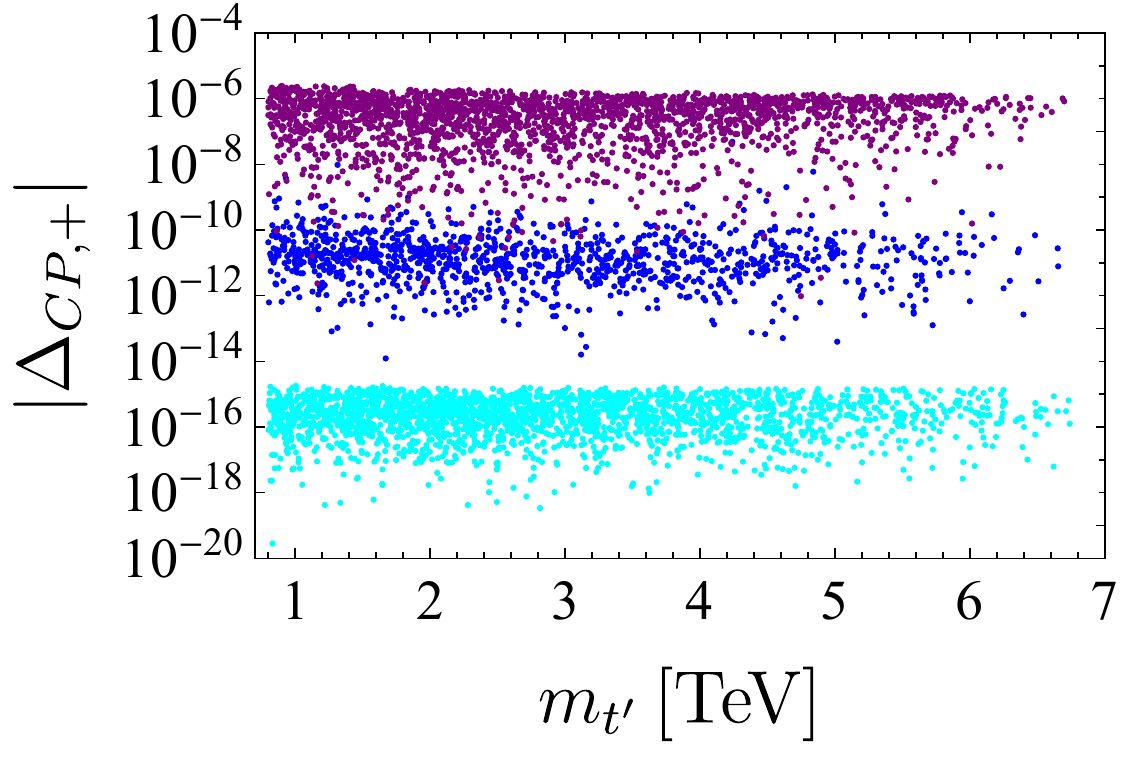}
         \includegraphics[width=0.49\textwidth]{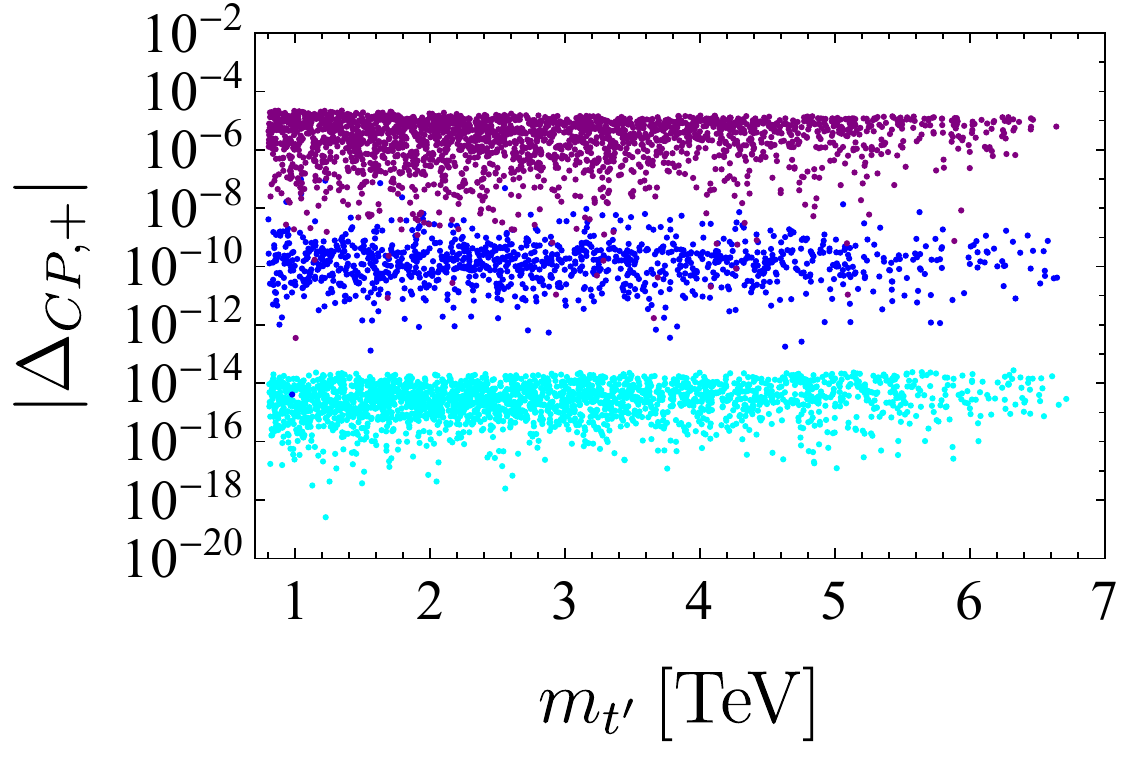}
         \includegraphics[width=0.49\textwidth]{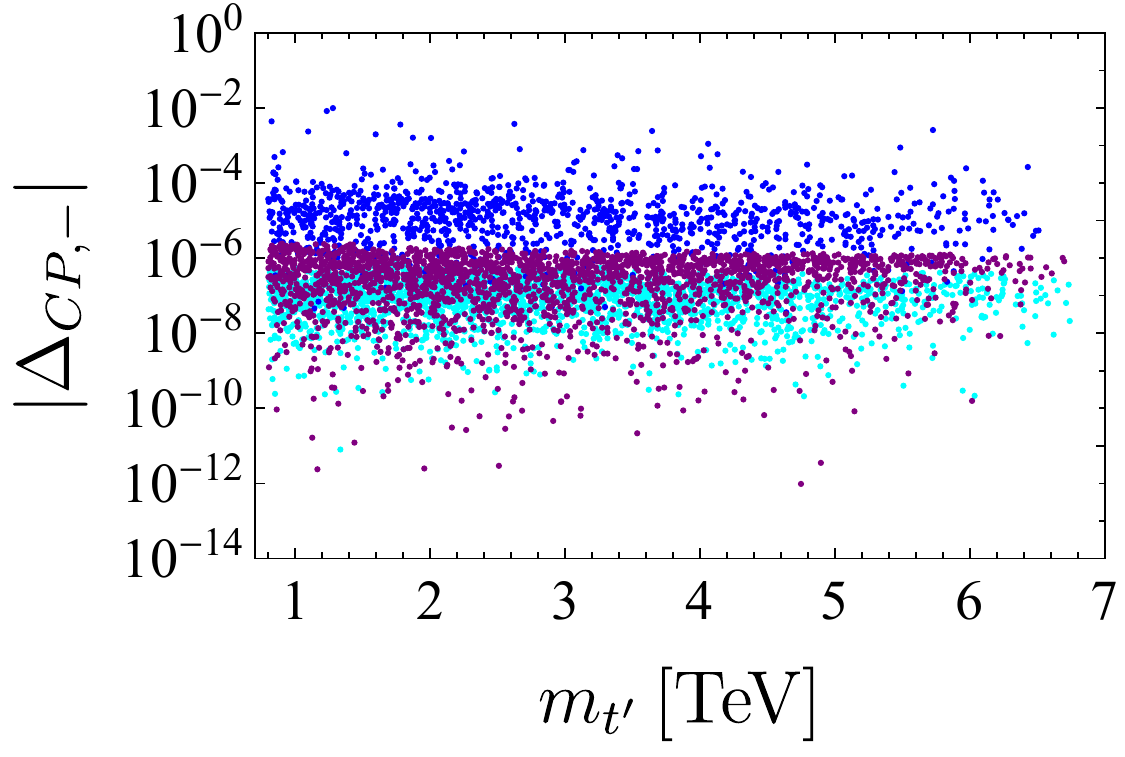}
         \includegraphics[width=0.46\textwidth]{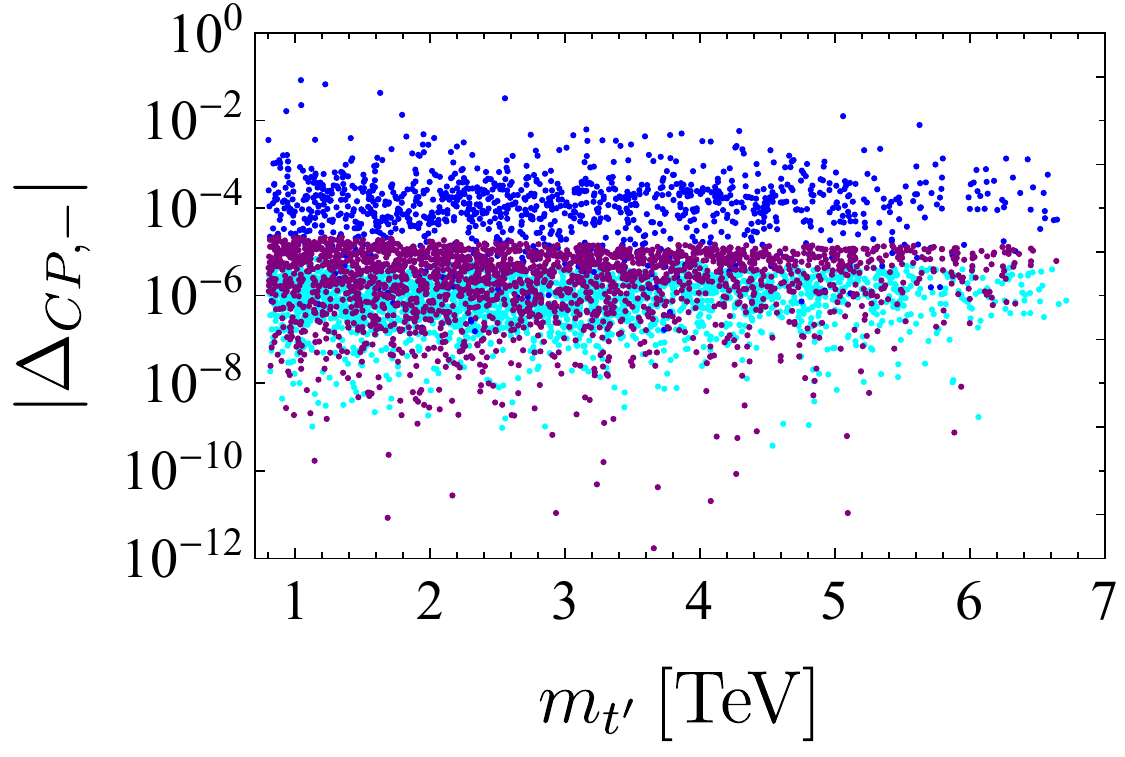}
    \caption{The branching ratio (top panels), the $|\Delta_{CP,+}|$ asymmetry (middle panels) and $|\Delta_{CP,-}|$ asymmetry (bottom panels) for the radiative decays of $t'$. The left (right) panels are for radiative decays with a final state photon (gluon). The cyan, blue and purple points correspond to $t'$ decay channels with a $u$, $c$ and $t$ quark final state respectively shown as a function of the vector--like quark mass $m_{t'}$ for the $3\sigma$ parameter space described in Ref.~\cite{Branco:2021vhs}.  }
    \label{fig:brcpscan}
\end{figure}

\paragraph{Full numerical scan}\mbox{}\\\\
It is instructive to perform a scan over the $3\sigma$ allowable regions for the mixing angles as well as the VLQ mass as shown in Figure 2 and Figure 3 of Ref.~\cite{Branco:2021vhs}. Implementing a numerical scan in this region and imposing the following $3\sigma$ constraints \footnote{It is important to note that the fit does not impose a unitarity requirement on the quark mixing matrix.}
\begin{align}
    |K_{\textrm{CKM}}|&=\begin{pmatrix}
    0.97370 + 0.00014 & 0.2245 + 0.0008 & (3.82 + 0.24)\times 10^{-3} \\
    0.221 \pm 0.004 & 0.987 \pm 0.011 & (41 \pm 1.4)\times10^{-3} \\
    (8 \pm 0.3)\times 10^{-3} & (38.1 \pm 1.1)\times 10^{-3} & 1.013 \pm 0.030
    \end{pmatrix},
\end{align}
on the $3\times3$ sub--matrix of $\mathcal{V}_L^\dagger$ which corresponds to the usual $V_\textrm{CKM}$. The results of the scan performed is shown in Figure~\ref{fig:brcpscan}.

Unsurprisingly, we see that that we get similar central values to the those around the best fit point shown in Figure~\ref{fig:brcp}. Studying the top left panel of Figure~\ref{fig:brcpscan} we see that we get the familiar branching ratio hierarchy where $\mathcal{B}(t'\rightarrow u \gamma)$ exceeds the $c\gamma$ and $t\gamma$ final state channels. We observer here, that the bulk of the points lie in the much larger range $10^{-12}\text{-}10^{-5}$ for the $m_{t'}$ values considered. We note that there is a larger allowable region for the decay into the second and third generation quarks because of the larger permitted range of quark mixing values in the scan for their respective generations. We find in the top right panel of Figure~\ref{fig:brcpscan}, that the branching ratios for the gluon channels fall in the approximate range $10^{-12}\text{-}10^{-5}$. We see, once again, that the branching ratios for decays into a gluon final state are larger than into a photon final state for the same reasons discussed in the previous section.

The absolute magnitude of $CP$ ratios are shown in the middle panels of Figure~\ref{fig:brcpscan} and we see that across all generations $|\Delta_{CP,+}(t'\rightarrow u_i\gamma)|\simeq 10^{-18}\text{-}10^{-6}$ and $|\Delta_{CP,+}(t'\rightarrow u_i g)|\simeq 10^{-18}\text{-}10^{-5}$  for $u_i=u,c,t$ quarks in the final state respectively. Similar to Figure~\ref{fig:brcp}, we see that the largest magnitudes for the photon and gluon channels correspond to the $t'\rightarrow t\gamma$ and $t'\rightarrow t g$ processes respectively. If we consider the largest $CP$ ratios shown in the bottom panels of Figure~\ref{fig:brcpscan}, we observe  $|\Delta_{CP,-}(t'\rightarrow u_i\gamma)|\simeq 10^{-12}\text{-}10^{-2}$ and $|\Delta_{CP,-}(t'\rightarrow u_i g)|\simeq 10^{-10}\text{-}10^{-1}$. The largest values being for both the photon and gluon channels being into the second generation, i.e. $t'\rightarrow c\gamma$ and $t'\rightarrow c g$ processes respectively. The full scan shows that very large and phenomenologically noteworthy $CP$ asymmetries from these VLQ decays can be realised. The CKM unitarity problem remains an important puzzle in particle physics and ongoing measurements in flavour sector will enable more precise determination of the size of these radiative decays and the $CP$ asymmetry. 

\subsection{\texorpdfstring{$CP$}{TEXT} violation with a vector isosinglet down--type quark}
\label{sec:CPdowntype}
In this section, we consider a model and global fit shown in Ref.~\cite{Alok:2014yua} where the SM is extended by the addition of a vector--like isosinglet down--type quark $b'$. The addition of a $b'$ leads to quark mixing via the $3\times4$ upper submatrix of the $4\times4$ quark mixing matrix $\mathcal{V}_L^\dagger$ shown in Appendix~\ref{sec:quarkmixingmatrix}.  We require the matrix $X^d=V^\dagger V$ to describe the FCN vertices with $h$, $\chi$ and $Z$ shown in Section~\ref{sec:lagrangian}. The addition of an isosinglet down--type quark $b'$ will modify the couplings of the SM bottom quark with $W$, $Z$ and Higgs boson. Deviations in these electroweak couplings, if measured, can provide indirect evidence of VLQs. The quark mixing matrix is parametrised according to Appendix~\ref{sec:quarkmixingmatrix}, with three SM mixing angles $\theta_{12}$, $\theta_{13}$ and $\theta_{23}$ and one $CP$ violating phase $\delta_{13}$ along with three new mixing angles $\theta_{14}, \theta_{24}$ and $\theta_{34}$ and three new $CP$ violating phases $\delta_{14}$ and $\delta_{24}$.

    In this fit, the authors consider constraints from several flavour observables \cite{Alok:2014yua}, they then compare the measured values of these processes with theoretical predictions from the SM and perform a $\chi^2$ fit and then redo this process with theoretical predictions with a $b'$ VLQ included and obtain values for the various mixing angles and phases mentioned above. This is a similar fit method to the one described in Section~\ref{sec:newsignalsuptype}. The results of the fit are shown in Table~\ref{tab:bp_bestfit} for benchmark masses $m_{b'}=800$ GeV and $m_{b'}=1200$ GeV.

\begin{table}[!h]
\begin{centering}
  \begin{tabular}{ |c | c | c | c | }
    \hline
    Parameter & SM &$m_{b'}=800$ GeV&$m_{b'}=1200$ GeV \\ \hline
    $\theta_{12}$ & $0.2273\pm 0.0007$ &  $0.2271\pm 0.0008$ & $0.2270\pm 0.0008$\\ 
    $\theta_{13}$ &  $0.0035\pm 0.0001$ &  $0.0038\pm 0.0001$&$0.0038\pm 0.0001$ \\
       $\theta_{23}$ & $0.039\pm 0.0007$ & $0.0391\pm 0.0007$&$0.0391\pm 0.0007$ \\
       $\delta_{13}$&$1.10\pm0.1$&$1.04\pm0.08$&$1.04\pm0.08$\\ \hline
       $\theta_{14}$& - &$0.0151 \pm 0.0154 $ &$0.0147 \pm 0.0149$\\
       $\theta_{24}$& - & $0.0031 \pm 0.0039 $&$0.0029 \pm 0.0036$\\
       $\theta_{34}$& - &$0.0133 \pm 0.0130$ &$0.0123 \pm 0.0122$\\
       $\delta_{14}$&-& $0.11 \pm 0.22$&$0.11 \pm 0.23$\\
       $\delta_{24}$&-&$3.23 \pm 0.24 $ &$3.23 \pm 0.27$\\\hline
       $\chi^2$/ d.o.f.&$71.15/60 $&$70.99/63 $&$70.96/63$\\
    \hline
  \end{tabular}
          \caption{Best fit parameters for the quark mixing matrix with a vector--like quark isosinglet $b'$ included. These are shown for two benchmark quark masses $m_{b'}=800~\rm GeV$ and $m_{b'}=1200~\rm GeV$ outlined in Table 4 of Ref.~\cite{Alok:2014yua}.}
    \label{tab:bp_bestfit}
\end{centering}
\end{table}

The resulting best fit values for the elements of the upper $3\times 3$ block of the quark mixing matrix are not significantly affected by the addition of $b'$ and are almost unchanged from the SM CKM fit parameters. Since we have the best fit values for the mixing matrix, we may now again apply the formalism similar to Section~\ref{sec:CPuptype} but this time for the down--type isosinglet $b'$ and compute its radiative decays into lighter down--type quarks in conjunction with a gluon or photon. We again compute branching ratios $\mathcal{B}$ and the $CP$ ratios $\Delta_{CP,+}$ and $\Delta_{CP,-}$. The results corresponding to each kinematically allowed decay are shown in Table~\ref{tab:vlqd_results}.

We find that the largest branching ratio for radiative decays of $b'$ for both $m_{b'}=800$ GeV and $m_{b'}=1200$ GeV is for the $b'\rightarrow bg$ process which has a relatively large $\mathcal{B}\simeq10^{-4}$ while decays to $dg$ and $sg$ final states are similar with $\mathcal{B}\simeq 10^{-8}\text{-}10^{-5}$. Like Section~\ref{sec:CPuptype}, we find decays to photon final states are in general about one order of magnitude smaller than to gluon final states. Similar to the previous sections, we find that the $CP$ ratios have a hierarchy $|\Delta_{CP,-}|>>|\Delta_{CP,+}|$  for all the $b'$ decays considered with the chosen fit. We get $|\Delta_{CP,+}|\simeq 10^{-13}\text{-}10^{-8}$ for a photon or gluon in the final state respectively. We observe much larger $CP$ asymmetries for $b'$ decays in the ratio $|\Delta_{CP,-}|\simeq 10^{-3}\text{-}10^{-1}$ for a photon or gluon in the final state. These are not dissimilar from the magnitudes of the $CP$ asymmetries shown in Section~\ref{sec:CPuptype}. We note that there will be no $CP$ asymmetry generated in correlated radiative decays such as $b\rightarrow s\gamma$ since the $b$--quark mass (and other light flavoured down--type quarks)  is too low to enable the threshold mass condition required to generate a complex phase in the loop functions.
\begin{table}[!ht]
    \centering
    \begin{tabular}{|c|c|c|c|} 
    \hline
   \multirow{2}{15mm}{Decay channel} & \multicolumn{3}{c|}{$m_{b'}=800~{\rm GeV}$} \\\cline{2-4}
  &   $\mathcal{B}\qquad$ & $\Delta_{CP,+}\qquad$ &  $\Delta_{CP,-}\qquad$ \\\hline
     $b'\to d \gamma$& $ (3.5^{+2.3}_{-3.5}\times10^{-6}$ &  $(-0.15^{+0.16}_{-2.60})\times10^{-12}$ &$(-0.33^{+0.3}_{-0.6})\times10^{-2}$  \\\hline
     $b'\to s \gamma$& $(0.007^{+4.0}_{-0.010})\times10^{-6}$ &$(2.1^{+8.0}_{-2.1})\times10^{-9}$ & $0.12^{+0.5}_{-0.18}$ \\\hline
     $b'\to b \gamma$&$ (1.0^{+1.3}_{-1.0})\times10^{-5} $ &$(0.14^{+5.00}_{-0.15})\times10^{-7} $ &$0.12^{+4.0}_{-0.13} \times10^{-2}$  \\\hline
     $b'\to d g$&$ (1.3^{+0.9}_{-1.3} )\times10^{-5} $ &$(-0.05^{+1.8}_{-1.2})\times10^{-11}$ &$-0.01^{+0.4}_{-0.27}$  \\\hline
     $b'\to s g$&$ (0.08^{+1.0}_{-0.008})\times10^{-6} $ &$(0.3^{+1.6}_{-1.7})\times10^{-8}$ &$  0.2^{+0.8}_{-1.0}$  \\\hline
     $b'\to b g$&$ 0.7^{+1.0}_{-0.7}\times10^{-4} $ &$(0.2^{+7.0}_{-0.4})\times10^{-7}$ &$(0.20^{+6.00}_{-0.34})\times10^{-2}$ \\\hline
    \end{tabular}
    \end{table}
    \begin{table}[!ht]
    \centering
    \begin{tabular}{|c|c|c|c|} 
    \hline
     \multirow{2}{15mm}{Decay channel} & \multicolumn{3}{c|}{$m_{b'}=1200~{\rm GeV}$} \\\cline{2-4}
  &   $\mathcal{B}\qquad$ & $\Delta_{CP,+}\qquad$ &  $\Delta_{CP,-}\qquad$ \\\hline
     $b'\to d \gamma$ &$ (2.6^{+1.7}_{-2.6})\times 10^{-6}$ &$(-0.10^{+0.10}_{-1.4}\times10^{-12})$ & $(-0.22^{+0.21}_{-2.80})\times10^{-2}$  \\\hline
     $b'\to s \gamma$&$ (0.003^{+4.0}_{-0.032})\times10^{-6}$ &$(0.25^{+1.2}_{-0.25}\times10^{-8}$ &$0.15^{+0.7}_{-0.18}$ \\\hline
     $b'\to b \gamma$ &$(6.0^{+9.0}_{-6.0})\times10^{-6} $ &$(0.10^{+8.0}_{-0.11})\times10^{-7}$ &$ (0.08^{+7.0}_{-0.09})\times 10^{-2}$ \\\hline
     $b'\to d g$&$(8.0^{+0.80}_{-1.2})\times10^{-6}$ &$(-0.05^{+0.028}_{-1.2})\times10^{-11}$ &$-0.011^{+0.008}_{-0.35}$ \\\hline
     $b'\to s g$&$(0.05^{+9.0}_{-0.05})\times10^{-6}$ &$(0.3^{+1.1}_{-1.5})\times10^{-8}$ &$(0.2^{+0.8}_{-1.2})$ \\\hline
     $b'\to b g$&$(5.0^{+8.0}_{-5.0})\times10^{-5}$ &$(0.17^{+5.0}_{-0.16})\times10^{-7}$ &$(0.16^{+6.0}_{-0.2})\times10^{-2}$ \\\hline
    \end{tabular}
    \caption{Results for the branching ratios and $CP$ asymmetries for the  radiative decay channels of the vector--like quark $b'$. These are based on the best fit mixing parameters and benchmark masses $m_{b'}=800~\rm GeV$ and $m_{b'}=1200~\rm GeV$ outlined in Table~\ref{tab:bp_bestfit}.}
    \label{tab:vlqd_results}    
    \end{table}

\subsection{Prospects for vector--like quark production at the LHC }
While the purpose of the analyses detailed above is to outline the theoretical prediction of various radiative decays and importantly the existence of nonzero $CP$ asymmetries induced in such decays. It is also important to consider discovery prospects through direct VLQ production at the LHC. The dominant contributions to the pair production cross section only depends on the mass of the VLQ in question. For $t'$ as considered in Section~\ref{sec:CPuptype}, at 13 TeV with $36.1\textrm{fb}^{-1}$ of data, we have cross sections exceeding $1$ fb for masses below $600$ GeV. However the cross section falls off quickly for larger masses, at $m_{t'}=1.4$ TeV we note that it is reduced to around $3\times10^{-3}\textrm{pb}$ \cite{ATLAS:2018ziw}. The details for the pair production cross section calculated at NNLO for different masses of the heavy $t'$ singlets is shown in Figure 2 of Ref.~\cite{ATLAS:2018ziw}.

If we take a benchmark point that would enable discovery of the $t'$ at the 95\% CL with mass $m_{t'}=1.2$ TeV, we observe a production cross section $\sigma(pp\to t'\bar{t}')\simeq 10^{-2}$pb which would produce $\mathcal{O}(10^2)$ events at $36.1\textrm{fb}^{-1}$. The largest decay channels considered in Section~\ref{sec:CPuptype} have a branching ratio of order $\mathcal{O}(10^{-5})$. Hence we may reasonably estimate that  we require at least a thousand times increase in data collected to probe the radiative decays and $CP$ asymmetries in these parameter regions. The HL-LHC expects to have a luminosity of approximately $3000 \textrm{fb}^{-1}$ \cite{Aberle:2749422} so this presents an important opportunity to search for such rare VLQ events. As discussed in Section\ref{sec:CPuptype}, correlated top quark decays may also be a way to search for the VLQ sector. Since the top quark production cross section at the $13$ TeV LHC is $\sigma(pp\to t\bar{t})\simeq 10^{3} \textrm{pb}$, at the luminosity $36.1\textrm{fb}^{-1}$, we expect $\mathcal{O}(10^{7})$ top quarks to be produced.  The maximum branching ratios including off shell $t'$ and FCN vertices is $\mathcal{O}(10^{-10})$ based on Table~\ref{tab:vlqu_results}. Thus we also require approximately a thousand times more data to be sensitive to radiative top decays.

For $b'$ as shown in Section~\ref{sec:CPdowntype}, we note that Figure 3(a) of Ref.~\cite{ATLAS:2018ziw} shows a very similar pair production cross section evolution to $t'$ with a maximum radiative branching ratio in Table~\ref{tab:vlqd_results} of $\mathcal{O}(10^{-5})$ as well. Hence the luminosity estimates required to discover $b'$ approximately coincide with those required for $t'$. Additionally, it should be noted that future prospective hadron colliders with higher centre--of--mass energies could increase the pair production cross section for VLQs and top quarks significantly leading to more rapid discovery or further exclusion of the VLQ sector.

\section{Conclusion}
The electromagnetic transition dipole moment of a vector--like quark isosinglet is computed analytically. The relationship between radiative decays induced by the off--diagonal components of the electromagnetic dipole moment and the chiral form factors are elucidated and the resulting branching ratios and corresponding $CP$ asymmetries are shown. These decays exist due to a combination of complex phases resulting from the loop functions and the extended quark mixing matrix and provide an exciting means to search for exotic quarks. We provide a full analytical formulation for the $CP$ asymmetry resulting from the loop functions as well as numerical studies for several global fits favouring inclusion of $t'$ and $b'$ vector isosinglets to the SM. In the fits considered for $t'$, we find phenomenological interesting results with branching ratios up to $\simeq10^{-5}$ and $CP$ asymmetries $\geq10^{-2}$. For $b'$, in the global fit considered, we find branching ratios $\simeq10^{-4}$ and $CP$ asymmetries up to order unity. Radiative decays of $b'$ in such a scenario would provide a tantalising probe for new physics and provide clean experimental signatures.Observation of rare vector--like quark radiative decays in the models studied would be expected at the HL--LHC with a factor of a thousand increase in data at current energies, or with even less at future colliders operating at higher centre--of--mass energy. The analytical results shown in this work can be applied to a variety of vector--like quark parameter scans and be easily extended to consider more exotic scenarios such as doublet and triplet vector--like quarks in the future.

\section*{Acknowledgements} 

SB would like to thank Maura Ramirez-Quezada and Ye-Ling Zhou for useful early discussions on vector--like quarks and helpful feedback that improved the clarity of the final text.

\appendix
\section{Matrix Elements and form factor relations}
\subsection{Lorentz invariant amplitudes}
\label{sec:feynmanamplitudes}
The matrix elements for the Feynman diagrams in Fig.~\ref{fig:SM_loops} are given by
\begin{align}
i\mathcal{M}_1&= \pm|Q_\alpha| A_\alpha \int \frac{d^4p}{(2\pi)^4}\frac{\overline{u}(p_\fin) \gamma_\mu\PL(\slashed{p}_\fin -\slashed{p}+m_\alpha)\gamma^\rho(\slashed{p}_\ini-\slashed{p}+m_\alpha)\gamma^\mu\PL u(p_\ini) \epsilon^*_\rho(q)}{[(p_\fin-p)^2-m_\alpha^2][p^2-\mW^2][(p_\ini-p)^2-m_\alpha^2]},\nonumber\\[.2cm]
i\mathcal{M}_2&= \pm |Q_\alpha| A_\alpha \int\frac{d^4p}{(2\pi)^4}\frac{\overline{u}(p_\fin)(m_\fin\PL-m_\alpha
    \PR)(\slashed{p}_\fin-\slashed{p}+m_\alpha)\gamma^\rho(\slashed{p}_\ini-\slashed{p}+m_\alpha)(m_\alpha\PL-\mnui\PR)u(p_\ini) \epsilon^*_\rho(q)}{m_W^2[(p_\fin-p)^2-m_\alpha^2][(p_\ini-p)^2-m_\alpha^2][p^2-\mW^2]},\nonumber\\[.2cm]
i\mathcal{M}_3&= A_\alpha \int \frac{d^4p}{(2\pi)^4}\frac{\overline{u}(p_\fin) \gamma_\nu\PL (\slashed{p}+m_\alpha)\gamma_\mu\PL V(p_\ini,p_\fin,p)^{\mu\nu\rho} u(p_\ini) \epsilon^*_\rho(q)}{[(p_\fin-p)^2-\mW^2][p^2-m_\alpha^2][(p_\ini-p)^2-\mW^2]},\nonumber\\[.2cm]
    i \mathcal{M}_4&= A_\alpha \int \frac{d^4p}{(2\pi)^4}\frac{\overline{u}(p_\fin)(\mnuj\PL-m_\alpha
    \PR)(\slashed{p}+m_\alpha)(m_\alpha\PL-\mnui\PR)(2p-p_\ini-p_\fin)^\rho u(p_\ini) \epsilon^*_\rho(q)}{m_W^2[(p_\fin-p)^2-\mW^2][p^2-m_\alpha^2][(p_\ini-p)^2-\mW^2]},\nonumber\\[.2cm]
   i\mathcal{M}_{5+6}&= A_\alpha \int \frac{d^4p}{(2\pi)^4}\overline{u}(p_\fin)\left[\frac{\gamma^\rho\PL (\slashed{p}+m_\alpha)(m_\alpha\PL-\mnui\PR)}{(p^2-m_\alpha^2)((p_\fin-p)^2-\mW^2)((p_\ini-p)^2-\mW^2)}\right.\nonumber\\[.2cm]
      &~~~~~~~~~~~~~~~~~~~~~~~~~~~~~~~~~~~~~~~~~~~\left.+\frac{(m_\alpha\PR-m_\fin\PL)
    (\slashed{p}+m_\alpha)\gamma^\rho\PL }{(p^2-m_\alpha^2)((p_\ini-p)^2-\mW^2)((p_\fin-p)^2-\mW^2)}\right]u(p_\ini) \epsilon^*_\rho(q)\nonumber,\\[.2cm]
          i   \mathcal{M}_7&= C_\beta \int \frac{d^4p}{(2\pi)^4}\frac{\overline{u}(p_\fin)(m_\fin\PL+    m_\beta\PR)(\slashed{p}_\fin-\slashed{p}+m_\beta)\gamma^\rho(\slashed{p}_\ini-\slashed{p}-m_\beta)(m_\beta\PL+
    m_\ini\PR) u(p_\ini) \epsilon^*_\rho(q)}{m_W^2[(p_\fin-p)^2-m_\beta^2][p^2-m_h^2][(p_\ini-p)^2-m_\beta^2]},\nonumber\\[.2cm]
    i\mathcal{M}_8&= -C_\beta \int \frac{d^4p}{(2\pi)^4}\frac{\overline{u}(p_\fin)(m_\fin\PL-
    m_\beta\PR)(\slashed{p}_\fin-\slashed{p}+m_\beta)\gamma^\rho(\slashed{p}_\ini-\slashed{p}-m_\beta)(m_\beta\PL-
    m_\ini\PR) u(p_\ini) \epsilon^*_\rho(q)}{m_W^2[(p_\fin-p)^2-m_\beta^2][p^2-m_Z^2][(p_\ini-p)^2-m_\beta^2]},\nonumber\\[.2cm]
         i\mathcal{M}_9&= -\int \frac{d^4p}{(2\pi)^4}\frac{\overline{u}(p_\fin)\gamma_\nu[ C^L_{\fin\beta}\PL+C^R_{\fin\beta}\PR](\slashed{p}_\fin-\slashed{p}+m_\beta)\gamma^\rho(\slashed{p}_\ini-\slashed{p}+m_\beta)\gamma^\nu[ C^L_{\beta \ini}\PL+C^R_{\beta \ini} \PR]  u(p_\ini) \epsilon^*_\rho(q)}{[(p_\fin-p)^2-m_\alpha^2][p^2-m_Z^2][(p_\ini-p)^2-m_\alpha^2]}.
\label{eq:loop1-6}
\end{align}
where the contribution from the triple gauge boson vertex in $\mathcal{M}_3$ is given by
\begin{eqnarray}
V^{\mu\nu\rho} &=& g^{\mu\nu}(2p_\ini-p-p_\fin)^\rho +
  g^{\rho\mu}(2p_\fin-p-p_\ini)^\nu +
  g^{\nu\rho}(2p-p_\ini-p_\fin)^\mu \,,
\end{eqnarray}
and the coefficient definitions given in \eqref{eq:uptypecoefficients} and \eqref{eq:downtypecoefficients} are utilised for an up-- and down--type quark in the initial state respectively. Technically, using the Feynman rules derived from Section~\ref{sec:lagrangian}, the amplitudes for a down--type quark in the initial state will have global minus applied to every amplitude in Eq.~\eqref{eq:loop1-6}. However, since this is a global minus sign, omitting this like we have above, will not affect any of the physics discussion presented in this work. Its important to note that the plus and minus branch of the first two amplitudes refers to photon and gluon emission respectively. 

\subsection{Form factor relations}
\label{sec:formfactorrelations}
Form factors relationships for the neutral vertex diagrams in terms of the analytical loop functions \eqref{eq:fcnloopfunctions}, are given by
\begin{footnotesize}
\begin{align}
   &{f^{\rm R}_{\fin\ini}}^{(7)}=\frac{C_\beta}{16\pi^2}K_{\fin\ini}^{(7)},&&{f^{\rm R}_{\ini\fin}}^{(7)}=\frac{C_\beta^*}{16\pi^2}K_{\ini\fin}^{(7)} ,&&{f^{\rm L}_{\fin\ini}}^{(7)}=\frac{C_\beta}{16\pi^2}K_{\ini\fin}^{(7)}, &&{f^{\rm L}_{\ini\fin}}^{(7)}=\frac{C_\beta^*}{16\pi^2}K_{\fin\ini}^{(7)},\nonumber\\
&{f^{\rm R}_{\fin\ini}}^{(8)}=-\frac{C_\beta}{16\pi^2}K_{\fin\ini}^{(8)},&&{f^{\rm R}_{\ini\fin}}^{(8)}=-\frac{C_\beta^*}{16\pi^2}K_{\ini\fin}^{(8)} ,&&{f^{\rm L}_{\fin\ini}}^{(8)}=-\frac{C_\beta}{16\pi^2}K_{\ini\fin}^{(8)}, &&{f^{\rm L}_{\ini\fin}}^{(8)}=-\frac{C_\beta^*}{16\pi^2}K_{\fin\ini}^{(8)},\nonumber\\
 &{f^{\rm R}_{\fin\ini}}^{(9LL)}=-\frac{C^L_{\fin\beta}C^L_{\beta\ini}}{16\pi^2}K_{\fin\ini}^{(9)},&&{f^{\rm R}_{\ini\fin}}^{(9LL)}=-\frac{{C^L_{\fin\beta}}^*{C^L_{\beta\ini}}^*}{16\pi^2}K_{\ini\fin}^{(9)}, &&{f^{\rm L}_{\fin\ini}}^{(9LL)}=-\frac{C^L_{\fin\beta}C^L_{\beta\ini}}{16\pi^2}K_{\ini\fin}^{(9)},&&{f^{\rm L}_{\fin\ini}}^{(9LL)}=-\frac{{C^L_{\fin\beta}}^*{C^L_{\beta\ini}}^*}{16\pi^2}K_{\fin\ini}^{(9)},\nonumber\\
 &{f^{\rm R}_{\fin\ini}}^{(9LR)}=-\frac{C^L_{\fin\beta}C^R_{\beta\ini}}{16\pi^2}K_{\fin\ini}^{(10)}, &&{f^{\rm R}_{\ini\fin}}^{(9LR)}=0, &&{f^{\rm L}_{\fin\ini}}^{(9LR)}=0, &&{f^{\rm L}_{\ini\fin}}^{(9LR)}=-\frac{{C^L_{\fin\beta}}^*{C^R_{\beta\ini}}^*}{16\pi^2}K_{\fin\ini}^{(10)},\nonumber\\
 &{f^{\rm R}_{\fin\ini}}^{(9RL)}=0,&&{f^{\rm R}_{\ini\fin}}^{(9RL)}=-\frac{{C^R_{\fin\beta}}^*{C^L_{\beta\ini}}^*}{16\pi^2}K_{\fin\ini}^{(10)},  &&{f^{\rm L}_{\fin\ini}}^{(9RL)}=-\frac{C^R_{\fin\beta}C^L_{\beta\ini}}{16\pi^2}K_{\fin\ini}^{(10)},&&{f^{\rm L}_{\ini\fin}}^{(9RL)}=0,\nonumber\\
&{f^{\rm R}_{\fin\ini}}^{(9RR)}=-\frac{C^R_{\fin\beta}C^R_{\beta\ini}}{16\pi^2}K_{\ini\fin}^{(9)},&&
 {{f^{\rm R}_{\ini\fin}}}^{(9RR)}=-\frac{{C^R_{\fin\beta}}^*{C^R_{\beta\ini}}^*}{16\pi^2}K_{\fin\ini}^{(9)},&&{f^{\rm L}_{\fin\ini}}^{(9RR)}=-\frac{C^R_{\fin\beta}C^R_{\beta\ini}}{16\pi^2}K_{\fin\ini}^{(9)},&&{f^{\rm L}_{\ini\fin}}^{(9RR)}=-\frac{{C^R_{\fin\beta}}^*{C^R_{\beta\ini}}^*}{16\pi^2}K_{\ini\fin}^{(9)},\label{eq:form_factors}
\end{align}
\end{footnotesize}
\section{Parametrisations of quark mixing matrix}
\subsection{Quark mixing matrix with isosinglet included}
\label{sec:quarkmixingmatrix}

It is convenient for us to define the generic $4 \times 4$ unitary quark mixing matrix $\mathcal{V}$. We define the Hermitian conjugate $\mathcal{V}_L^\dagger$ in terms of three SM mixing angles $\theta_{12}$, $\theta_{13}$ and $\theta_{23}$ and $CP$ violating phase $\delta_{13}$. However it also required three new angles $\theta_{14}, \theta_{24}$ and $\theta_{34}$ along with three new $CP$ violating phases $\delta_{14}$ and $\delta_{24}$.

\begin{align}
\label{eq:CKM4}
    \mathcal{V}_L^\dagger&=\begin{pmatrix}
1 & 0 & 0 & 0\\
0 & 1 & 0 & 0\\
0 & 0 & c_{34} & s_{34} \\
0 & 0 & -s_{34} & c_{34}
\end{pmatrix} 
\begin{pmatrix}
1 & 0 & 0 & 0\\
0 & c_{24} & 0 & s_{24}e^{-i\delta_{24}}\\
0 & 0 &1 & 0 \\
0 & -s_{24}e^{i\delta_{24}} & 0 & c_{24}
\end{pmatrix}
\begin{pmatrix}
c_{14} & 0 & 0 & s_{14}e^{-i\delta_{14}}\\
0 & 1 & 0 & 0\\
0 & 0 & 1 & 0 \\
-s_{14}e^{i\delta_{14}}  & 0 & 0 & c_{14}
\end{pmatrix} \nonumber \\ 
&
\begin{pmatrix}
1 & 0 & 0 & 0\\
0 & c_{23} & s_{23} & 0\\
0 & -s_{23} & c_{23} & 0 \\
0 & 0 & 0 & 1
\end{pmatrix} 
\begin{pmatrix}
c_{13} & 0 & s_{13}e^{-i\delta_{13}} & 0\\
0 & 1 & 0 & 0\\
-s_{13}e^{i\delta_{13}} & 0 & c_{13} & 0 \\
0 & 0 & 0 & 1
\end{pmatrix} 
\begin{pmatrix}
c_{12} & s_{12} & 0 & 0\\
-s_{12} & c_{12} & 0 & 0\\
0 & 0 & 1 & 0 \\
0 & 0 & 0 & 1
\end{pmatrix} 
\end{align}
where $c_{ij}=\cos\theta_{ij}$ and $s_{ij}=\sin\theta_{ij}$. We can extract the left--handed mixing matrix $V$ required for \eqref{eq:LagrangianW} from $\mathcal{V}_L^\dagger$. In the case of a $t'$ isosinglet, such as in Section~\ref{sec:CPuptype}, $V$ is the $4\times3$ sub--matrix of $\mathcal{V}_L^\dagger$. Analogously, in the case of a $b'$ isosinglet as in Section~\ref{sec:CPdowntype}, $V$ refers to the $3\times4$ sub--matrix of $\mathcal{V}_L^\dagger$. In both cases, if we consider the limit $\theta_{14}, \theta_{24}, \theta_{34} \rightarrow 0$, there is no mixing between the SM quarks and the new isosinglet quarks and $V$ reduces to the CKM matrix.
\subsection{Wolfenstein parametrisation of extended mixing matrix}
\label{sec:Wolfenstein}
If we consider a $t'$ VLQ only, $V$ is the left--hand $4\times3$ submatrix of the $4\times4$ unitary $\mathcal{V}_L^\dagger$ in Eq.~\eqref{eq:CKM4}. Here it is best to choose a parametrisation of $\mathcal{V}_L^\dagger$ such that the new matrix elements $V_{41}$, $V_{42}$ and $V_{43}$ take simple forms. Using the Hou--Soni--Steger parametrisation we have the elements of $V$ given by
\begin{align} 
V_{12} &= \lambda \,, &V_{23}&=A\lambda^2 \,, &V_{13}&= A\lambda^3 C e^{i\delta_{13}}\nonumber \\
V_{41} &= -P\lambda^3 e^{i\delta_{41}} \,, &V_{42}&=-Q\lambda^2 e^{i\delta_{42}} \,, &V_{43} &= -r\lambda.  
\end{align}
There are four SM parameters $\lambda$, $A$, $C$, $\delta_{13}$ and five new physics parameters $P$, $Q$, $r$, $\delta_{41}$, $\delta_{42}$. Of the remaining six CKM matrix elements, $V_{11}$, $V_{21}$ and $V_{22}$ retain their SM parametrisations 
\begin{align}
    V_{11}&=1-\frac{\lambda^2}{2}\,, &V_{21}&=-\lambda\,, &V_{22}=1-\frac{\lambda^2}{2}
\end{align}
and the third row is given by
\begin{align} 
V_{31} &= A\lambda^3(1-Ce^{i\delta_{13}})-P r \lambda^4  e^{i\delta_{41}}+\frac{1}{2}AC\lambda^5e^{i\delta_{13}}, \nonumber \\
V_{32} &=-A \lambda^2 - Q r \lambda^3 e^{i\delta_{42}}+A\lambda^4 \left(\frac{1}{2}-C e^{i\delta_{13}}\right), &V_{33} =& 1 - \frac{1}{2}r^2\lambda^2.
\end{align}
In the limit $P,Q,r\rightarrow 0$, there is no mixing between the SM quarks and $t'$ and $V$ reduces to the standard Wolfenstein parametrisation of the CKM matrix \cite{Wolfenstein:1983yz}.

\bibliographystyle{JHEP} 
\bibliography{refs_vlq}

\providecommand{\href}[2]{#2}\begingroup\raggedright\begin{thebibliography}{10}

\bibitem{Deshpande:1981zq}
N.~G. Deshpande and G.~Eilam, \emph{{FLAVOR CHANGING ELECTROMAGNETIC
  TRANSITIONS}}, \href{http://dx.doi.org/10.1103/PhysRevD.26.2463}{\emph{Phys.
  Rev. D} {\bf 26} (1982) 2463}.

\bibitem{Beneke:2000hk}
M.~Beneke et~al., \emph{{Top quark physics}},  in \emph{{Workshop on Standard
  Model Physics (and more) at the LHC (First Plenary Meeting)}}, pp.~419--529,
  3, 2000.
\newblock \href{http://arxiv.org/abs/hep-ph/0003033}{{\tt hep-ph/0003033}}.

\bibitem{Balaji:2019fxd}
S.~Balaji, M.~Ramirez-Quezada and Y.-L. Zhou, \emph{{CP violation and circular
  polarisation in neutrino radiative decay}},
  \href{http://dx.doi.org/10.1007/JHEP04(2020)178}{\emph{JHEP} {\bf 04} (2020)
  178}, [\href{http://arxiv.org/abs/1910.08558}{{\tt 1910.08558}}].

\bibitem{Balaji:2020fxd}
S.~Balaji, M.~Ramirez-Quezada and Y.-L. Zhou, \emph{{CP violation in the
  neutrino dipole moment}},  \href{http://arxiv.org/abs/2008.12795}{{\tt
  2008.12795}}.

\bibitem{Balaji:2020qjg}
S.~Balaji, \emph{{$CP$ asymmetries in the rare top decays $t\to c\gamma$ and
  $t\to c g$}},
  \href{http://dx.doi.org/10.1103/PhysRevD.102.113010}{\emph{Phys. Rev. D} {\bf
  102} (2020) 113010}, [\href{http://arxiv.org/abs/2009.03315}{{\tt
  2009.03315}}].

\bibitem{Dedes:2014asa}
A.~Dedes, M.~Paraskevas, J.~Rosiek, K.~Suxho and K.~Tamvakis, \emph{{Rare
  Top-quark Decays to Higgs boson in MSSM}},
  \href{http://dx.doi.org/10.1007/JHEP11(2014)137}{\emph{JHEP} {\bf 11} (2014)
  137}, [\href{http://arxiv.org/abs/1409.6546}{{\tt 1409.6546}}].

\bibitem{Hill:1994hp}
C.~T. Hill, \emph{{Topcolor assisted technicolor}},
  \href{http://dx.doi.org/10.1016/0370-2693(94)01660-5}{\emph{Phys. Lett. B}
  {\bf 345} (1995) 483--489}, [\href{http://arxiv.org/abs/hep-ph/9411426}{{\tt
  hep-ph/9411426}}].

\bibitem{Gaitan:2015hga}
R.~Gaitán, J.~H. Montes~de Oca, E.~A. Garcés and R.~Martinez, \emph{{Rare top
  decay $t \rightarrow c \gamma$ with flavor changing neutral scalar
  interactions in two Higgs doublet model}},
  \href{http://dx.doi.org/10.1103/PhysRevD.94.094038}{\emph{Phys. Rev. D} {\bf
  94} (2016) 094038}, [\href{http://arxiv.org/abs/1503.04391}{{\tt
  1503.04391}}].

\bibitem{Atwood:1996vj}
D.~Atwood, L.~Reina and A.~Soni, \emph{{Phenomenology of two Higgs doublet
  models with flavor changing neutral currents}},
  \href{http://dx.doi.org/10.1103/PhysRevD.55.3156}{\emph{Phys. Rev. D} {\bf
  55} (1997) 3156--3176}, [\href{http://arxiv.org/abs/hep-ph/9609279}{{\tt
  hep-ph/9609279}}].

\bibitem{Arhrib:2005nx}
A.~Arhrib, \emph{{Top and Higgs flavor changing neutral couplings in two Higgs
  doublets model}},
  \href{http://dx.doi.org/10.1103/PhysRevD.72.075016}{\emph{Phys. Rev. D} {\bf
  72} (2005) 075016}, [\href{http://arxiv.org/abs/hep-ph/0510107}{{\tt
  hep-ph/0510107}}].

\bibitem{DiazCruz:1989ub}
J.~Diaz-Cruz, R.~Martinez, M.~Perez and A.~Rosado, \emph{{Flavor Changing
  Radiative Decay of Thf T Quark}},
  \href{http://dx.doi.org/10.1103/PhysRevD.41.891}{\emph{Phys. Rev. D} {\bf 41}
  (1990) 891--894}.

\bibitem{Atwood:1995ud}
D.~Atwood, L.~Reina and A.~Soni, \emph{{Probing flavor changing top - charm -
  scalar interactions in $e^{+} e^{-}$ collisions}},
  \href{http://dx.doi.org/10.1103/PhysRevD.53.1199}{\emph{Phys. Rev. D} {\bf
  53} (1996) 1199--1201}, [\href{http://arxiv.org/abs/hep-ph/9506243}{{\tt
  hep-ph/9506243}}].

\bibitem{Atwood:1995ej}
D.~Atwood, L.~Reina and A.~Soni, \emph{{Flavor changing neutral scalar currents
  at $\mu^{+} \mu^{-}$ colliders}},
  \href{http://dx.doi.org/10.1103/PhysRevLett.75.3800}{\emph{Phys. Rev. Lett.}
  {\bf 75} (1995) 3800--3803}, [\href{http://arxiv.org/abs/hep-ph/9507416}{{\tt
  hep-ph/9507416}}].

\bibitem{Balaji:2018zna}
S.~Balaji, R.~Foot and M.~A. Schmidt, \emph{{Chiral SU(4) explanation of the
  $b\to s$ anomalies}},
  \href{http://dx.doi.org/10.1103/PhysRevD.99.015029}{\emph{Phys. Rev. D} {\bf
  99} (2019) 015029}, [\href{http://arxiv.org/abs/1809.07562}{{\tt
  1809.07562}}].

\bibitem{Balaji:2019kwe}
S.~Balaji and M.~A. Schmidt, \emph{{Unified SU(4) theory for the $R_{D^{(*)}}$
  and $R_{K^{(*)}}$ anomalies}},
  \href{http://dx.doi.org/10.1103/PhysRevD.101.015026}{\emph{Phys. Rev. D} {\bf
  101} (2020) 015026}, [\href{http://arxiv.org/abs/1911.08873}{{\tt
  1911.08873}}].

\bibitem{Grinstein:2010ve}
B.~Grinstein, M.~Redi and G.~Villadoro, \emph{{Low Scale Flavor Gauge
  Symmetries}}, \href{http://dx.doi.org/10.1007/JHEP11(2010)067}{\emph{JHEP}
  {\bf 11} (2010) 067}, [\href{http://arxiv.org/abs/1009.2049}{{\tt
  1009.2049}}].

\bibitem{Cacciapaglia:2010vn}
G.~Cacciapaglia, A.~Deandrea, D.~Harada and Y.~Okada, \emph{{Bounds and Decays
  of New Heavy Vector-like Top Partners}},
  \href{http://dx.doi.org/10.1007/JHEP11(2010)159}{\emph{JHEP} {\bf 11} (2010)
  159}, [\href{http://arxiv.org/abs/1007.2933}{{\tt 1007.2933}}].

\bibitem{ATLAS:2018ziw}
{\scshape ATLAS} collaboration, M.~Aaboud et~al., \emph{{Combination of the
  searches for pair-produced vector-like partners of the third-generation
  quarks at $\sqrt{s} =$ 13 TeV with the ATLAS detector}},
  \href{http://dx.doi.org/10.1103/PhysRevLett.121.211801}{\emph{Phys. Rev.
  Lett.} {\bf 121} (2018) 211801}, [\href{http://arxiv.org/abs/1808.02343}{{\tt
  1808.02343}}].

\bibitem{Boehm:2017nrl}
C.~B\oe~hm, C.~Degrande, O.~Mattelaer and A.~C. Vincent, \emph{{Circular
  polarisation: a new probe of dark matter and neutrinos in the sky}},
  \href{http://dx.doi.org/10.1088/1475-7516/2017/05/043}{\emph{JCAP} {\bf 05}
  (2017) 043}, [\href{http://arxiv.org/abs/1701.02754}{{\tt 1701.02754}}].

\bibitem{Eilam:1990zc}
G.~Eilam, J.~Hewett and A.~Soni, \emph{{Rare decays of the top quark in the
  standard and two Higgs doublet models}},
  \href{http://dx.doi.org/10.1103/PhysRevD.44.1473}{\emph{Phys. Rev. D} {\bf
  44} (1991) 1473--1484}.

\bibitem{AguilarSaavedra:2002ns}
J.~Aguilar-Saavedra and B.~Nobre, \emph{{Rare top decays t ---> c gamma, t --->
  cg and CKM unitarity}},
  \href{http://dx.doi.org/10.1016/S0370-2693(02)03230-6}{\emph{Phys. Lett. B}
  {\bf 553} (2003) 251--260}, [\href{http://arxiv.org/abs/hep-ph/0210360}{{\tt
  hep-ph/0210360}}].

\bibitem{AguilarSaavedra:2002kr}
J.~Aguilar-Saavedra, \emph{{Effects of mixing with quark singlets}},
  \href{http://dx.doi.org/10.1103/PhysRevD.69.099901}{\emph{Phys. Rev. D} {\bf
  67} (2003) 035003}, [\href{http://arxiv.org/abs/hep-ph/0210112}{{\tt
  hep-ph/0210112}}].

\bibitem{Alok:2015vvk}
A.~K. Alok, S.~Banerjee, D.~Kumar, S.~U. Sankar and D.~London,
  \emph{{New-physics signals of a model with an isosinglet vector-like t'
  quark}}, \href{http://dx.doi.org/10.22323/1.234.0579}{\emph{PoS} {\bf
  EPS-HEP2015} (2015) 579}.

\bibitem{Belfatto:2019swo}
B.~Belfatto, R.~Beradze and Z.~Berezhiani, \emph{{The CKM unitarity problem: A
  trace of new physics at the TeV scale?}},
  \href{http://dx.doi.org/10.1140/epjc/s10052-020-7691-6}{\emph{Eur. Phys. J.
  C} {\bf 80} (2020) 149}, [\href{http://arxiv.org/abs/1906.02714}{{\tt
  1906.02714}}].

\bibitem{Belfatto:2021jhf}
B.~Belfatto and Z.~Berezhiani, \emph{{Are the CKM anomalies induced by
  vector-like quarks? Limits from flavor changing and Standard Model precision
  tests}}, \href{http://dx.doi.org/10.1007/JHEP10(2021)079}{\emph{JHEP} {\bf
  10} (2021) 079}, [\href{http://arxiv.org/abs/2103.05549}{{\tt 2103.05549}}].

\bibitem{Alok:2014yua}
A.~K. Alok, S.~Banerjee, D.~Kumar and S.~Uma~Sankar, \emph{{Flavor signatures
  of isosinglet vector-like down quark model}},
  \href{http://dx.doi.org/10.1016/j.nuclphysb.2016.03.012}{\emph{Nucl. Phys. B}
  {\bf 906} (2016) 321--341}, [\href{http://arxiv.org/abs/1402.1023}{{\tt
  1402.1023}}].

\bibitem{Alok:2015iha}
A.~K. Alok, S.~Banerjee, D.~Kumar, S.~U. Sankar and D.~London,
  \emph{{New-physics signals of a model with a vector-singlet up-type quark}},
  \href{http://dx.doi.org/10.1103/PhysRevD.92.013002}{\emph{Phys. Rev. D} {\bf
  92} (2015) 013002}, [\href{http://arxiv.org/abs/1504.00517}{{\tt
  1504.00517}}].

\bibitem{Hou:1987hm}
W.-S. Hou, A.~Soni and H.~Steger, \emph{{Effects of a Fourth Family on $b \to s
  \gamma$ and a Useful Parametrization of Quark Mixing for Rare $B$ Decays}},
  \href{http://dx.doi.org/10.1016/0370-2693(87)90135-3}{\emph{Phys. Lett. B}
  {\bf 192} (1987) 441}.

\bibitem{Zyla:2020zbs}
{\scshape Particle Data Group} collaboration, P.~Zyla et~al., \emph{{Review of
  Particle Physics}}, \href{http://dx.doi.org/10.1093/ptep/ptaa104}{\emph{PTEP}
  {\bf 2020} (2020) 083C01}.

\bibitem{Chetyrkin:2000yt}
K.~G. Chetyrkin, J.~H. Kuhn and M.~Steinhauser, \emph{{RunDec: A Mathematica
  package for running and decoupling of the strong coupling and quark masses}},
  \href{http://dx.doi.org/10.1016/S0010-4655(00)00155-7}{\emph{Comput. Phys.
  Commun.} {\bf 133} (2000) 43--65},
  [\href{http://arxiv.org/abs/hep-ph/0004189}{{\tt hep-ph/0004189}}].

\bibitem{Herren:2017osy}
F.~Herren and M.~Steinhauser, \emph{{Version 3 of RunDec and CRunDec}},
  \href{http://dx.doi.org/10.1016/j.cpc.2017.11.014}{\emph{Comput. Phys.
  Commun.} {\bf 224} (2018) 333--345},
  [\href{http://arxiv.org/abs/1703.03751}{{\tt 1703.03751}}].

\bibitem{Branco:2021vhs}
G.~C. Branco, J.~T. Penedo, P.~M.~F. Pereira, M.~N. Rebelo and J.~I.
  Silva-Marcos, \emph{{Addressing the CKM unitarity problem with a vector-like
  up quark}}, \href{http://dx.doi.org/10.1007/JHEP07(2021)099}{\emph{JHEP} {\bf
  07} (2021) 099}, [\href{http://arxiv.org/abs/2103.13409}{{\tt 2103.13409}}].

\bibitem{DeSimone:2012fs}
A.~De~Simone, O.~Matsedonskyi, R.~Rattazzi and A.~Wulzer, \emph{{A First Top
  Partner Hunter's Guide}},
  \href{http://dx.doi.org/10.1007/JHEP04(2013)004}{\emph{JHEP} {\bf 04} (2013)
  004}, [\href{http://arxiv.org/abs/1211.5663}{{\tt 1211.5663}}].

\bibitem{Buchkremer_2013}
M.~Buchkremer, G.~Cacciapaglia, A.~Deandrea and L.~Panizzi,
  \emph{Model-independent framework for searches of top partners},
  \href{http://dx.doi.org/10.1016/j.nuclphysb.2013.08.010}{\emph{Nuclear
  Physics B} {\bf 876} (Nov, 2013) 376–417}.

\bibitem{Botella_2017}
F.~J. Botella, G.~C. Branco, M.~Nebot, M.~N. Rebelo and J.~I. Silva-Marcos,
  \emph{Vector-like quarks at the origin of light quark masses and mixing},
  \href{http://dx.doi.org/10.1140/epjc/s10052-017-4933-3}{\emph{The European
  Physical Journal C} {\bf 77} (Jun, 2017) }.

\bibitem{Aaboud_2018}
M.~Aaboud, G.~Aad, B.~Abbott, O.~Abdinov, B.~Abeloos, D.~Abhayasinghe et~al.,
  \emph{Combination of the searches for pair-produced vectorlike partners of
  the third-generation quarks at s=13 tev with the atlas detector},
  \href{http://dx.doi.org/10.1103/physrevlett.121.211801}{\emph{Physical Review
  Letters} {\bf 121} (Nov, 2018) }.

\bibitem{Sirunyan_2019}
A.~Sirunyan, A.~Tumasyan, W.~Adam, F.~Ambrogi, T.~Bergauer, J.~Brandstetter
  et~al., \emph{Search for pair production of vectorlike quarks in the fully
  hadronic final state},
  \href{http://dx.doi.org/10.1103/physrevd.100.072001}{\emph{Physical Review D}
  {\bf 100} (Oct, 2019) }.

\bibitem{Aberle:2749422}
O.~Aberle, \emph{{High-Luminosity Large Hadron Collider (HL-LHC): Technical
  design report}}.
\newblock CERN Yellow Reports: Monographs. CERN, Geneva, 2020,
  \href{http://dx.doi.org/10.23731/CYRM-2020-0010}{10.23731/CYRM-2020-0010}.

\bibitem{Wolfenstein:1983yz}
L.~Wolfenstein, \emph{{Parametrization of the Kobayashi-Maskawa Matrix}},
  \href{http://dx.doi.org/10.1103/PhysRevLett.51.1945}{\emph{Phys. Rev. Lett.}
  {\bf 51} (1983) 1945}.

\end{thebibliography}\endgroup
\end{document}